\documentclass[12pts]{article}
\usepackage{setspace}
\usepackage{amsfonts}
\usepackage{amsmath}
\usepackage{amssymb}
\usepackage{amsthm}
\usepackage{authblk}
\usepackage{hyperref}
\usepackage{fullpage}
\usepackage{graphicx}
\usepackage{geometry}
\usepackage{multirow}
\usepackage{multicol}
\usepackage{xcolor}
\usepackage{caption}
\usepackage{subfigure}
\newtheorem{theorem}{Theorem}[section]

\newtheorem{definition}{Definition}
\title{Parametric Analysis of Bivariate Current Status data with Competing risks using Frailty model}
\author[1]{Biswadeep Ghosh}
\author[1]{Anup Dewanji}
\author[2]{Sudipta Das}
\affil[1]{Applied Statistics Unit, Indian Statistical Institute}
\affil[2]{Ramakrishna Mission Vivekananda Educational and Research Institute}
\date{}

\begin{document}
	

\maketitle

{\bf Abstract.} Shared and correlated Gamma frailty models are widely used in the literature to model the association in multivariate current status data. In this paper, we have proposed two other new Gamma frailty models, namely shared cause-specific and correlated cause-specific Gamma frailty to capture association in bivariate current status data with competing risks. We have investigated identifiability of the bivariate models with competing risks for each of the four frailty variables. We have considered maximum likelihood estimation of the model parameters. Thorough simulation studies have been performed to study finite sample behaviour of the estimated parameters. Also, we have analyzed a real data set on hearing loss in two ears using Exponential type and Weibull type cause-specific baseline hazard functions with the four different Gamma frailty variables and compare the fits using AIC. 

\section{Introduction}

In current status data, a special case of interval censored data, whether  failure has occurred or not is observed at a monitoring or observation time. In such situations, if the failure can be due to one of several causes, then we have current status data with competing risks. If we have current status data on failure of multiple individuals in a cluster who may be related, or failure of multiple organs or components from a single individual, and each failure is subject to several competing risks, then this is known as multivariate current status data with competing risks. In this paper, we focus on bivariate current status data with competing risks. Although such bivariate data may arise from two different scenarios, as discussed above, we henceforth refer to the first one with two individuals being in a cluster or pair. 
For example, we may have current status data on incidence time of first major disease from a couple (husband and wife) while the different major diseases may be treated as the competing risks. On the other hand, we have current status data on hearing loss in both right and left ears in which the hearing loss may be due to three different primary reasons, namely, SNHL, Conductive and Mixed (See Koley and Dewanji, 2022).  \\

Dependence between the two failure times can be modeled through bivariate cumulative hazard vector (Johnson and Kotz, 1975; Dabrowska, 1988). Recently, there are two popular approaches, one using copula (Jewell et al., $2005$; Wang et al., $2008$; Ma et al., $2015$;
Prenen et al., $2017$; Hu et al., $2017$;
Cui et al., $2018$; Sun et al., $2021$) and the other through introduction of random frailty (Duchateau, $2008$ and Wienke, $2010$). For bivariate failure time with competing risks, there have been some work using bivariate cumulative hazard vector (See Ansa and Sankaran, 2005; Sankaran et al., 2006). Modeling multivariate failure time with competing risks using frailty seems to be natural (See Gorfine and Hsu, $2011$) because of its simple and appealing interpretation, while the same using copula does not seem to be that straightforward. \\

Although there is some work on analysis of multivariate failure time data with competing risks (See Ansa and 
Sankaran, $2005$; Sankaran et al., $2006$; Gorfine and Hsu, $2011$), we do not know of any work on bivariate current status data with competing risks. 
In this paper, we consider analysis of bivariate current status data with competing risks while the modeling of dependence uses the popular Gamma frailty. We consider four different types of Gamma frailty variables depending on the association at individual and cause-specific level (namely, Shared, Correlated, Shared cause-specific, Correlated cause-specific, respectively) to capture different type of associations in the presence of competing risks (See Gorfine and Hsu, $2011$).  \\

We consider a parametric class of cause-specific baseline hazard function described in the next section. Since identifiability is an important issue while working with frailty variable, we also investigate identifiability of each of the four models corresponding to the four different frailty variables, as stated above. For each model, we derive the popular "cross-hazard ratio" measure (Bandeen-Roche and Liang, 2002) to study the nature of association. Two of our proposed frailty models allow for dependence at the level of competing risks, unlike the commonly used shared and correlated frailty models, incorporating a variety of dependency structures. \\

In Section $2$, we describe the bivariate current status data and the four models with proofs of the corresponding identifiability results. We also compute the cross-hazard ratio as a function of the bivariate failure time $(t_1,t_2)$ for all possible pairs of causes. In Section $3$, we discuss maximum likelihood estimation of the associated parameters. We also study the asymptotic properties of maximum likelihood estimators (MLEs) of the parameters under some regularity conditions. Section $4$ presents a detailed simulation study to investigate finite sample properties of the MLEs. Section $5$ considers analysis of the hearing loss data using the four models and compares them using the AIC criterion. Section 6 ends with  some concluding remarks. \\

\section{The Models and Identifiability}

Let $T_1,T_2$ be the random variables denoting the failure time of the first and second individual of a pair, respectively, with $J_1,J_2$ denoting the corresponding causes of failure. We assume that $J_k$ takes values in the set $\{1,2,\cdots,L_{k}\}$, for $k = 1,2$, implying that the two individuals may be subject to different sets of competing risks. In some of the models though, we need to consider the same set of competing risks for the two individuals. \\ 

Note that we do not observe the quadruplet $(T_1,T_2,J_1,J_2)$ directly for the bivariate current status data with competing risks. Rather we observe the failure status at some monitoring time. If the observed status for an individual is `failure', then we also observe the cause of failure $J_1$ or $J_2$. 
Let $X_1,X_2$ be random variables denoting the monitoring times for the first and the second individual, respectively. This allows for different monitoring time for the two individuals although, in most situations, the two individuals have the same monitoring time denoted by a random variable $X$. We assume that the monitoring times are independent of the individual failure times without involving any common parameter. Hence, our observed data is the quadruplet $(X_1,X_2,J_1,J_2)$, with the convention that $J_1$ or $J_2$ is equal to 0 when the corresponding individual's status at the monitoring time  is `not a failure', or censored. \\

The sub-distribution function of the $k$th individual due to cause $j$  at time $t_{k}$ is 
$$F^{(k)}_{j}\big(t_k;\boldsymbol{\eta}\big) =
P\Big[T_{k} \leq t_{k},J_{k} = j;\boldsymbol{\eta}\Big] =  \int_{0}^{t_{k}}f^{(k)}_{j}\big(u;\boldsymbol{\eta}\big)du, $$
for all $t_{k} > 0,\ j = 1,\cdots,L_{k}$ and $k = 1,2$, where 
$f^{(k)}_{j}\big(u;\boldsymbol{\eta}\big)$ is the sub-density function for the $k$th individual due to cause $j$ and $\boldsymbol{\eta}$ is the associated parameter vector. The  cumulative distribution function for the $k$th individual at time 
$t_{k}$ is, therefore, 
$$F^{(k)}(t_{k};\boldsymbol{\eta}) = P\Big[T_{k} \leq t_{k};\boldsymbol{\eta}\Big] = 
\sum\limits_{j = 1}^{L_{k}}F^{(k)}_j\big(t_{k};\boldsymbol{\eta}\big), $$ 
so that the survival function for the $k$th individual is 
$$S^{(k)}(t_k;\boldsymbol{\eta}) = P\big[T_{k} > t_{k};\boldsymbol{\eta}\big] =1- F^{(k)}(t_{k};\boldsymbol{\eta}).$$

In our context of bivariate failure time, or current status, data with competing risks, the primary quantity of interest, however, is the joint sub-distribution function given by
$$F_{j_{1}j_{2}}(t_1,t_2;\boldsymbol{\eta}) = P\big[T_{1} \leq t_1,T_{2} \leq t_2,J_{1} = j_{1},J_{2} = j_2;\boldsymbol{\eta}\big],$$
for $t_{k} > 0, j_{k} = 1,\cdots,L_{k}$ and $k = 1,2$. Explicit expressions for all these quantities require modeling of dependence between the failure times of two individuals, possibly differently for different failure types. As discussed in Section 1, we model this dependence by means of different types of frailty variables, in particular, having some Gamma distributions. In our development of the model, in addition to the frailty distributions, we also require modeling of the cause-specific baseline hazard functions, as described  below. \\

The $j$th cause-specific hazard function of the $k$th individual at time $t_k$ is defined as 
$$\lambda^{(k)}_{j}\big(t_{k};\boldsymbol{\eta}\big) = \lim\limits_{h \to 0+}\frac{P\Big[t_{k} < T_{k} < t_{k} + h,J_{k} = j;\boldsymbol{\eta} | T_{k} \geq t_{k}\Big]}{h} = \frac{f^{(k)}_{j}\big(t_k;\boldsymbol{\eta}\big)}{S^{(k)}\big(t_{k};\boldsymbol{\eta}\big)},$$
for all $t_{k} > 0,\ j=1,\cdots,L_k$ and $k = 1,2$. \\

As already discussed, the dependence in bivariate failure time data with competing risks is modeled with the help of frailty variables. In general, let $\boldsymbol{\epsilon^{(k)}} = (\epsilon^{(k)}_1,\cdots,\epsilon^{(k)}_{L_k})$ denote the vector of frailty variables corresponding to different failure types for the $k$th individual, for $k = 1,2$. Note that the parameter vector $\boldsymbol{\eta}$ consists of two sets of parameters, one for the baseline cause-specific hazard functions, denoted by $\boldsymbol{\xi}$, and the other for the frailty models, denoted by $\boldsymbol{\theta}$, so that $\boldsymbol{\eta}=(\boldsymbol{\xi},\boldsymbol{\theta})$. 
We consider the general multiplicative model for the conditional cause-specific hazard function of the $k$th individual, given the frailty vector $\boldsymbol{\epsilon^{(k)}}$, as 
\begin{equation}\label{EqGeneralmodel}
    \lambda^{(k)}_{j}(t_k;\boldsymbol{\xi}|\boldsymbol{\epsilon^{(k)}}) = h^{(k)}_{0j}(t_k;\boldsymbol{\xi})\epsilon^{(k)}_{j},
\end{equation}
for $t_{k} > 0,\ j = 1,\cdots,L_{k}$ and $k = 1,2$, where  $h^{(k)}_{0j}(t_k;\boldsymbol{\xi})$ is the $j$th 
cause-specific baseline hazard function for the $k$th individual. The corresponding cumulative cause-specific baseline hazard function for the $k$th individual is 
$H^{(k)}_{0j}(t_k;\boldsymbol{\xi}) = \int\limits_{0}^{t_k}h^{(k)}_{0j}(u;\boldsymbol{\xi})du$.\\

As already mentioned, we consider parametric models for the cause-specific baseline hazard functions $h^{(k)}_{0j}(t_k;\boldsymbol{\xi})$'s with associated parameter vector  $\boldsymbol{\xi} \in \boldsymbol{\Xi}$. We assume these parametric cause-specific baseline hazard functions  $h^{(k)}_{0j}(t_k;\boldsymbol{\xi})$, for $j=1,\cdots,L_k,\ k=1,2$,  to be identifiable within $\boldsymbol{\Xi}$ in the sense that if, for some $\boldsymbol{\xi}$ and $\boldsymbol{\tilde{\xi}}$ in $\boldsymbol{\Xi}$, the equality $h^{(k)}_{0j}(t_k;\boldsymbol{\xi}) = h^{(k)}_{0j}(t_k;\boldsymbol{\tilde{\xi}}),$ 
    for all $t_k > 0,\ j=1,\cdots,L_k,$ and $k=1,2$, implies $\boldsymbol{\xi}=\boldsymbol{\tilde{\xi}}$. \\
    
For our purpose, we introduce a new parametric class of cause-specific baseline hazard functions $h^{(k)}_{0j}(t_k;\boldsymbol{\xi})$'s which includes expressions of $Exponential$, $Weibull$, $Gamma$, $Log-logistic$ and $Weibull-Gompertz$ distributions as special cases. This parametric class is particularly helpful as it contains various types of cause-specific baseline hazard functions such as increasing, decreasing and bathtub-shaped. We define this parametric class of $h^{(k)}_{0j}(t_k;\boldsymbol{\xi})$, suppressing the dependence on $k$ and $j$ for simplicity, as  
\begin{equation}\label{parametric_family_haz}
    h(t;\gamma,\alpha) = a(\gamma,\alpha)t^{\gamma - 1}b(t;\gamma,\alpha),
\end{equation} 
for all $t > 0$, where $\alpha>0$ and $\gamma>0$ are scale and shape parameters, respectively, $a(\gamma,\alpha)$ is a positive-valued function of $\gamma$ and $\alpha$; it is also assumed that   $a(\gamma,\alpha)$ is an one-one function in $\alpha$ for fixed $\gamma$. The function $b(t;\gamma,\alpha)$ is positive and such that $\lim\limits_{t \to 0+}b(t;\gamma,\alpha) = 1$. It can be easily proved, letting $\lim\limits_{t \to 0+}$, that this parametric class (\ref{parametric_family_haz}) is identifiable in $\alpha$ and $\gamma$ in the above-mentioned sense. 
Note that the parameters $\alpha$ and $\gamma$ depend on $k$ and $j$ in general. Therefore, the parameter $\boldsymbol{\xi}$ is the vector of all these $\alpha$'s and $\gamma$'s. The special cases are given below. \\

\begin{enumerate}
    \item With $\gamma = 1, a(1,\alpha)=\alpha$ and $b(t,1,\alpha) = 1$ for all $t > 0$, we have the $Exponential$ hazard with $h(t;\gamma,\alpha) = \alpha$, a constant. 
    \item With $a(\gamma,\alpha)=\gamma \alpha^{\gamma}$ and $b(t,\gamma,\alpha) = 1$, for all $t > 0$, we have $h(t;\gamma,\alpha) = \{\alpha\}^{\gamma}\gamma t^{\gamma - 1}$, the $Weibull$ hazard. 
    \item With $a(\gamma,\alpha) = \frac{\alpha^{\gamma}}{\Gamma{(\gamma)}}$ and $b(t,\gamma,\alpha) = \frac{\Gamma{(\gamma)}\exp{\{-t\alpha\}}}{\int\limits_{t\alpha}^{\infty}e^{-y}y^{\gamma - 1}dy}$, for all $t > 0$, we have the $Gamma$ hazard of the form $h(t;\gamma,\alpha) = \{\alpha\}^{\gamma}\frac{t^{\gamma - 1}\exp{\{-t\alpha\}}}{\int\limits_{t\alpha}^{\infty}e^{-y}y^{\gamma - 1}dy}$. 
    \item With $a(\gamma,\alpha) = \gamma \alpha^{\gamma}$ and $b(t,\gamma,\alpha) = \frac{1}{1 + (t\alpha)^{\gamma}}$, for all $t > 0$, we have the $Log-logistic$ hazard given by  
    $h(t;\gamma,\alpha) = \frac{\alpha^{\gamma}\gamma t^{\gamma - 1}}{1 + (t\alpha)^{\gamma}}$, which is a non-monotonic (upside bathtub shaped) function.
    \item  With $a(\gamma,\alpha) = \gamma \alpha^{\gamma}$ and $b(t,\gamma,\alpha) = e^{\alpha t}$ for all $t > 0$, we have the expression similar to that of  \textit{Weibull-Gompertz} hazard (dos Santos et al., 1995) given by $h(t;\gamma,\alpha) = \gamma \alpha^{\gamma} t^{\gamma - 1}e^{ \alpha t}$, which is a non-monotonic function.
\end{enumerate}

The conditional survival function of $k$th individual, given $\boldsymbol{\epsilon^{(k)}}$, is 
\begin{align*}
    S^{(k)}(t_k;\boldsymbol{\xi}|\boldsymbol{\epsilon^{(k)}}) &= \exp{\bigg[-\int\limits_{0}^{t_k}\sum\limits_{j=1}^{L_k}\lambda^{(k)}_{j}(u_k;\boldsymbol{\xi}|\boldsymbol{\epsilon^{(k)}})du_k\bigg]}\\
    &= \exp{\bigg[-\sum\limits_{j=1}^{L_k}\int\limits_{0}^{t_k}h^{(k)}_{0j}(u_k;\boldsymbol{\xi})\epsilon^{(k)}_{j}du_k\bigg]}\\
    &= \exp{\bigg[-\sum\limits_{j=1}^{L_k} \epsilon^{(k)}_{j} 
    H^{(k)}_{0j}(t_k;\boldsymbol{\xi})\bigg]}
\end{align*}
for $t_{k} > 0,j = 1,\cdots,L_{k}$ and $k = 1,2$. 
The conditional $j$th sub-distribution function of the $k$th individual, given $\boldsymbol{\epsilon^{(k)}}$, is 
\begin{align*}
    F^{(k)}_{j}(t_{k};\boldsymbol{\xi}|\boldsymbol{\epsilon^{(k)}}) &= \int\limits_{0}^{t_k}\lambda^{(k)}_{j}(u_k;\boldsymbol{\xi}|\boldsymbol{\epsilon^{(k)}})S^{(k)}(u_k|\boldsymbol{\epsilon^{(k)}})du_k\\
    &= \int\limits_{0}^{t_k}h^{(k)}_{0j}(u_k;\boldsymbol{\xi})\epsilon^{(k)}_{j}\exp{\bigg[-\sum\limits_{i=1}^{L_k}H^{(k)}_{0i}(u_k;\boldsymbol{\xi})\epsilon^{(k)}_{i}\bigg]}du_k,
\end{align*}
for $t_{k} > 0,\ j = 1,\cdots,L_{k}$ and $k = 1,2$.\\

Using the conditional independence between $T_1$ and $T_2$, given the frailty vectors $\boldsymbol{\epsilon^{(1)}}$ and $\boldsymbol{\epsilon^{(2)}}$, the unconditional joint sub-distribution function $F_{j_{1}j_{2}}\big(t_1,t_2;\boldsymbol{\xi},\boldsymbol{\theta}\big)$ 
at time $t_1$ and $t_2$ corresponding to causes $j_1$ and $j_2$, respectively, is 
\begin{align*}
& \mathbb{E}\bigg[F^{(1)}_{j_1}(t_{1};\boldsymbol{\xi}|\boldsymbol{\epsilon^{(1)}})F^{(2)}_{j_2}(t_{2};\boldsymbol{\xi}|\boldsymbol{\epsilon^{(2)}})\bigg]\\
&= \mathbb{E}\Bigg[\int\limits_{0}^{t_1}\int\limits_{0}^{t_2} 
\prod_{k=1}^2\left(
h^{(k)}_{0j_k}(u_k;\boldsymbol{\xi}) \epsilon^{(k)}_{j_k} \exp{\bigg(-
\sum\limits_{j'=1}^{L_k}H^{(k)}_{0j'}(u_k;\boldsymbol{\xi})\epsilon^{(k)}_{j'}\bigg)}\right)du_2du_1\Bigg]\\
&= \int\limits_{0}^{\infty}\cdots\int\limits_{0}^{\infty}\Bigg[\int\limits_{0}^{t_1}\int\limits_{0}^{t_2} \prod_{k=1}^2\left(
h^{(k)}_{0j_k}(u_k;\boldsymbol{\xi}) \epsilon^{(k)}_{j_k} \exp{\bigg(-\sum\limits_{j'=1}^{L_k}H^{(k)}_{0j'}(u_k;\boldsymbol{\xi})\epsilon^{(k)}_{j'}\bigg)}\right)du_2du_1\Bigg] g(\boldsymbol{\epsilon^{(1)}},\boldsymbol{\epsilon^{(2)}};\boldsymbol{\theta})d\boldsymbol{\epsilon^{(1)}}d\boldsymbol{\epsilon^{(2)}}, 
\end{align*}
for $t_{k} > 0,\ j_{k} = 1,\cdots,L_{k}$ and $k = 1,2$, where  $g(\boldsymbol{\epsilon^{(1)}},\boldsymbol{\epsilon^{(2)}};\boldsymbol{\theta})$ denotes the joint density of the frailty vectors $\boldsymbol{\epsilon^{(1)}}$ and $\boldsymbol{\epsilon^{(2)}}$ with the associated parameter vector $\boldsymbol{\theta}\in \boldsymbol{\Theta}$. In the following sub-sections, we shall consider different types of Gamma density for this frailty vector $(\boldsymbol{\epsilon^{(1)}},\boldsymbol{\epsilon^{(2)}})$. \\

The corresponding unconditional joint sub-density function  $f_{j_{1}j_{2}}\big(t_1,t_2;\boldsymbol{\xi},\boldsymbol{\theta}\big)$ is  
$$\int\limits_{0}^{\infty}\cdots\int\limits_{0}^{\infty}\Bigg[\prod_{k=1}^2\left(
h^{(k)}_{0j_k}(t_k;\boldsymbol{\xi}) \epsilon^{(k)}_{j_k} \exp{\bigg(-
\sum\limits_{j'=1}^{L_k}H^{(k)}_{0j'}(t_k;\boldsymbol{\xi})\epsilon^{(k)}_{j'}\bigg)}\right)\Bigg] g(\boldsymbol{\epsilon^{(1)}},\boldsymbol{\epsilon^{(2)}};\boldsymbol{\theta})d\boldsymbol{\epsilon^{(1)}}d\boldsymbol{\epsilon^{(2)}}.$$

It is of interest to investigate identifiability of the model for bivariate failure time with competing risks $(T_1,T_2,J_1,J_2)$ given by the joint sub-distribution function 
$F_{j_{1}j_{2}}\big(t_1,t_2;\boldsymbol{\xi},\boldsymbol{\theta}\big)$ with different Gamma frailty distributions. Note that these  identifiability results are important while analyzing bivariate current status data with competing risks $(X_1,X_2,J_1,J_2)$ since the likelihood contributions (See Section 3) can be obtained from the joint sub-distribution function. \\

\begin{definition}
   The model (\ref{EqGeneralmodel}) for bivariate failure time with competing risks is identifiable 
   within $\boldsymbol{\Xi}\times\boldsymbol{\Theta}$ 
   if, for some $\boldsymbol{\xi},\ \boldsymbol{\tilde{\xi}}\in\ \boldsymbol{\Xi}$ and $\boldsymbol{\theta},\ \boldsymbol{\tilde{\theta}}\in\ \boldsymbol{\Theta}$, the equality 
$$F_{j_{1}j_{2}}\big(t_1,t_2;\boldsymbol{\xi},\boldsymbol{\theta}\big) = F_{j_{1}j_{2}}\big(t_1,t_2;\boldsymbol{\tilde{\xi}},\boldsymbol{\tilde{\theta}}\big), $$
for all $t_{k} > 0,\ j_{k} = 1,\cdots,L_{k}$ and $k=1,2$, implies $\boldsymbol{\xi}=\boldsymbol{\tilde{\xi}}$ and $\boldsymbol{\theta}=\boldsymbol{\tilde{\theta}}$. 
\end{definition}

In the context of modeling bivariate failure time, `conditional hazard ratio' proposed by Oakes $(1989)$ is a popular measure of dependence. Bandeen-Roche and Liang $(2002)$ extended this measure for bivariate failure time with competing risks and named it `cross-ratio function'. 
This cross-Ratio function, denoted by $CR_{j_1,j_2}(t_1,t_2)$, between the failure of first individual at time $t_1$ due to cause $j_1$ and the failure of second individual at time $t_2$ due to cause $j_2$ is defined as 
$$CR_{j_{1}j_{2}}(t_1,t_2;\boldsymbol{\xi},\boldsymbol{\theta}) = \frac{S(t_1,t_2;\boldsymbol{\xi},\boldsymbol{\theta})f_{j_{1}j_{2}}(t_1,t_2;\boldsymbol{\xi},\boldsymbol{\theta})}
{\left(\int\limits_{t_1}^{\infty}\sum\limits_{j = 1}^{L_1}f_{jj_{2}}(u_1,t_2;\boldsymbol{\xi},\boldsymbol{\theta})du_1\right) \times \left(\int\limits_{t_2}^{\infty}\sum\limits_{j = 1}^{L_2}f_{j_{1}j}(t_1,u_2;\boldsymbol{\xi},\boldsymbol{\theta})du_2\right)}$$
for $t_{k} > 0,\ j_k = 1,\cdots,L_k$ and $k=1,2$. 
It is easy to see that this cross-ratio function is strictly greater than $0$. If $CR_{j_{1}j_{2}}(t_1,t_2;\boldsymbol{\xi},\boldsymbol{\theta})=1$, there is no association between failure of the first individual at time $t_1$ due to cause $j_1$ and failure of the second individual at time $t_2$ due to cause $j_2$. When $CR_{j_{1}j_{2}}(t_1,t_2;\boldsymbol{\xi},\boldsymbol{\theta})$ is larger (smaller) than 1, we have  positive (negative) association between such failures of the two individuals. In the following sub-sections, in addition to investigating model identifiability, we also derive this cross-ratio function for the models arising from the four different Gamma frailty distributions. 

\subsection{Shared Gamma frailty model}

The shared frailty model considers a common frailty variable for both the individuals regardless of their failure types. Dependence between the two failure times and types is induced by this shared frailty. As a result, the model can accommodate two different sets of competing risks. The model is given by   
\begin{equation}\label{shared_Gamma_frailty_model}
\lambda^{(k)}_{j}\big(t_k;\boldsymbol{\xi}| \epsilon\big) = h^{(k)}_{0j}\big(t_k;\boldsymbol{\xi}\big)\epsilon,  
\end{equation}
for $t_{k} > 0,\ j = 1,\cdots,L_{k}$ and $k = 1,2$, where $\epsilon$ is the shared frailty variable. \\

We assume the shared frailty $\epsilon$ to follow $\text{Gamma}(\frac{1}{\sigma^2},\frac{1}{\sigma^2})$ distribution, for some $\sigma > 0$, with density given by 
$$g(\epsilon;\sigma) = \frac{1}{\sigma^{\frac{2}{\sigma^2}}\Gamma(\frac{1}{\sigma^2})}e^{-\frac{\epsilon}{\sigma^2}}\epsilon^{\frac{1}{\sigma^2} - 1},$$
having mean 1 and variance $\sigma^2$. The frailty parameter space is, therefore, given by  $\boldsymbol{\Theta} = \{\sigma: \sigma > 0\}$. Shared Gamma frailty model, for example, may be used for cancer occurrence times of father and son, with the competing risks referring to different sites, because there is a common genetic and environmental effect shared by both of them. \\

Following the discussion in the beginning of this section, for this shared Gamma frailty model, conditional on the shared frailty $\epsilon$, we have  
$$S^{(k)}(t_{k};\boldsymbol{\xi}|\epsilon) = \exp{\Big[-\sum\limits_{j=1}^{L_k}H^{(k)}_{0j}(t_k;\boldsymbol{\xi})\epsilon\Big]}= \exp{\Big[- \epsilon   H^{(k)}_{0}(t_k;\boldsymbol{\xi})\Big]},$$
where $H^{(k)}_{0}(t_k;\boldsymbol{\xi})=\sum\limits_{j=1}^{L_k}H^{(k)}_{0j}(t_k;\boldsymbol{\xi})$, and 
$$F^{(k)}_{j}(t_k;\boldsymbol{\xi}|\epsilon) = \int\limits_{0}^{t_k}h^{(k)}_{0j}(u_k;\boldsymbol{\xi})\epsilon\exp{\Big[- \epsilon H^{(k)}_{0}(u_k;\boldsymbol{\xi})\Big]}du,$$
for $t_{k} > 0,\ j=1,\cdots,L_k$ and $k = 1,2$. Therefore, 
the unconditional joint sub-distribution function can be obtained, using conditional independence and then integrating with respect to the Gamma density of $\epsilon$, as 
$$F_{j_{1}j_{2}}(t_1,t_2;\boldsymbol{\xi},\sigma) = \int\limits_{0}^{t_1}\int\limits_{0}^{t_2}\frac{(1 + \sigma^2) \left(\prod_{k=1}^2 h^{(k)}_{0,j_k}(u_k;\boldsymbol{\xi})\right) du_2du_1}{\left[1 + \sigma^{2}\sum_{k=1}^2 
H^{(k)}_{0}(u_k;\boldsymbol{\xi})\right]^{2 + \frac{1}{\sigma^2}}},$$
with the corresponding  unconditional joint sub-density function as 
\begin{center}
$\displaystyle{f_{j_{1}j_{2}}(t_1,t_2;\boldsymbol{\xi},\sigma) = \frac{\partial^{2}}{\partial t_1 \partial t_2}F_{j_{1}j_{2}}(t_1,t_2;\boldsymbol{\xi},\sigma)
= \frac{(1 + \sigma^2) \prod_{k=1}^2 
h^{(k)}_{0j_k}(t_k;\boldsymbol{\xi})}{\left[1 + \sigma^2
\sum_{k=1}^2 H^{(k)}_{0}(t_k;\boldsymbol{\xi})
\right]^{2 + \frac{1}{\sigma^2}}}},$
\end{center}
for all $t_{k} > 0,\  j_k = 1,\cdots,L_{k}$ and $k = 1,2$. 
The  unconditional joint survival function is obtained as 
$$S(t_1,t_2;\boldsymbol{\xi},\sigma) = \int_0^{\infty}\left( \prod_{k=1}^2 S^{(k)}(t_k;\boldsymbol{\xi}|\epsilon)\right) g(\epsilon;\sigma) d\epsilon = 
\Big[1 + \sigma^2\sum_{k=1}^2 H^{(k)}_{0}(t_k;\boldsymbol{\xi})\Big]^{- \frac{1}{\sigma^2}},$$
for all $t_1,t_2 > 0$. \\

For the shared frailty model, as per Definition 2, we have the following definition of identifiability. 
\begin{definition}
The shared Gamma frailty model $(\ref{shared_Gamma_frailty_model})$ is identifiable within  $\boldsymbol{\Xi}\times\boldsymbol{\Theta}$ with 
$\boldsymbol{\Theta} = \{\sigma: \sigma > 0\}$ if, for some $\boldsymbol{\xi},\tilde{\boldsymbol{\xi}} \in \boldsymbol{\Xi}$ and $\sigma,\tilde{\sigma} \in \boldsymbol{\Theta}$, the equality of 
$F_{j_{1}j_{2}}(t_1,t_2;\boldsymbol{\xi},\sigma)$ and 
$F_{j_{1}j_{2}}(t_1,t_2;\tilde{\boldsymbol{\xi}},\tilde{\sigma}^2)$ with the above expressions, for all $t_{k} > 0,\ j_{k} = 1,\cdots,L_{k}$ and $k = 1,2$, implies $\boldsymbol{\xi} =\tilde{\boldsymbol{\xi}}$ and $\sigma = \tilde{\sigma}$. 
\end{definition}

Note that, for the parametric family of cause-specific  baseline hazard functions described in (\ref{parametric_family_haz}), the parameter vector $\boldsymbol{\xi}$ can be written as 
$$\boldsymbol{\xi}=\{(\gamma^{(k)}_j,\alpha^{(k)}_j),\ j=1,\cdots,L_k,\ k=1,2\}.$$
We can also write $\boldsymbol{\xi}= (\boldsymbol{\xi^{(1)}},\boldsymbol{\xi^{(2)}})$, where $\boldsymbol{\xi^{(k)}}=\{(\gamma^{(k)}_j,\alpha^{(k)}_j),\ j=1,\cdots,L_k\}$, for $k=1,2$. With this break-up of notation, it is to be noted that the unconditional $j$th sub-distribution function for the $k$th individual depends only on $\boldsymbol{\xi^{(k)}}$ and $\sigma$ and is given by 
\begin{equation}\label{marg_subdis_shared_Gamma_frailty}
    F^{(k)}_{j}(t_k;\boldsymbol{\xi^{(k)}},\sigma) =\int\limits_{0}^{t_k} \frac{ a(\gamma^{(k)}_{j},\alpha^{(k)}_{j})u^{\gamma^{(k)}_{j} - 1}b(u;\gamma^{(k)}_{j},\alpha^{(k)}_{j})du}{\big[1 + \sigma^{2} H^{(k)}_{0}(u;\boldsymbol{\xi^{(k)}})\big]^{1 + \frac{1}{\sigma^2}}},
\end{equation}
for all $t_k>0,\ j = 1,\cdots,L_{k}$ and $k = 1,2.$

\begin{theorem}
The shared Gamma frailty model (\ref{shared_Gamma_frailty_model}),  with the cause-specific baseline hazard functions belonging to the family defined in (\ref{parametric_family_haz}), is identifiable within $\boldsymbol{\Xi}\times\boldsymbol{\Theta}$.
\end{theorem}

\begin{proof}
The equality of the joint unconditional sub-distribution functions with $(\boldsymbol{\xi},\sigma)$ and $(\boldsymbol{\tilde{\xi}},\tilde{\sigma})$ implies equality of the corresponding sub-distribution functions $F^{(k)}_{j}(t_k;\boldsymbol{\xi^{(k)}},\sigma)$ and $F^{(k)}_{j}(t_k;\boldsymbol{\tilde{\xi}^{(k)}},\tilde{\sigma})$, for all $t_k>0,\ j = 1,\cdots,L_{k}$ and $k = 1,2.$ 
Differentiating both sides with respect to $t_k$, for fixed $k$,  we get equality of the corresponding unconditional sub-density functions to have 
\begin{equation*}
   \frac{ a(\gamma^{(k)}_{j},\alpha^{(k)}_{j})t^{\gamma^{(k)}_{j} - 1}_{k}b(t_k;\gamma^{(k)}_{j},\alpha^{(k)}_{j})}{\left[1 + \sigma^{2} H^{(k)}_{0}(t_k;\boldsymbol{\xi^{(k)}})\right]^{1 + \frac{1}{\sigma^2}}} =  \frac{ a(\tilde{\gamma}^{(k)}_{j},\tilde{\alpha}^{(k)}_{j})t^{\tilde{\gamma}^{(k)}_{j} - 1}_{k}b(t_k;\tilde{\gamma}^{(k)}_{j},\tilde{\alpha}^{(k)}_{j})} {\left[1 + \tilde{\sigma}^2 H^{(k)}_{0}(t_k;\boldsymbol{\tilde{\xi}^{(k)}})\right]^{1 + \frac{1}{\tilde{\sigma}^{2}}}},
\end{equation*}
for $k=1,2$. After rearranging the terms, we get 
\begin{equation}\label{identi_Eq_shared_Gamma_frailty_model}
   1 = t_k^{\tilde{\gamma}^{(k)}_{j} - \gamma^{(k)}_{j}}\frac{a(\tilde{\gamma}^{(k)}_{j},\tilde{\alpha}^{(k)}_{j})}{a(\gamma^{(k)}_{j},\alpha^{(k)}_{j})} \times    \frac{b(t_k;\tilde{\gamma}^{(k)}_{j},\tilde{\alpha}^{(k)}_{j}) \left[1 + \sigma^{2} H^{(k)}_{0}(t_k;\boldsymbol{\xi^{(k)}})\right]^{1 + \frac{1}{\sigma^2}}}{b(t_k;\gamma^{(k)}_{j},\alpha^{(k)}_{j})\left[1 + \tilde{\sigma}^2 H^{(k)}_{0}(t_k;\boldsymbol{\tilde{\xi}^{(k)}})\right]^{1 + \frac{1}{\tilde{\sigma}^2}}}.
\end{equation}
Now, letting $t_{k} \to 0+$, the right hand side of \ref{identi_Eq_shared_Gamma_frailty_model} tends to 0 or $\infty$ depending on whether $\tilde{\gamma}^{(k)}_{j}$ is larger or smaller than $\gamma^{(k)}_{j}$, while the left hand side remains at 1, resulting in a contradiction. Hence, $\tilde{\gamma}^{(k)}_{j}=\gamma^{(k)}_{j}$ for all $j=1,\cdots,L_k$ and $k=1,2$. Heckman and Singer ($1984$) used a similar technique for univariate failure time model without competing risks. Now, putting this equality in (\ref{identi_Eq_shared_Gamma_frailty_model}) and again letting $t_{k} \to 0+$, we get 
 $\tilde{\alpha}^{(k)}_{j}=\alpha^{(k)}_{j}$ for all $j=1,\cdots,L_k$ and $k=1,2$, since $a(\gamma,\alpha)$ is a one-one function in $\alpha$ for fixed $\gamma$.  So, we have $\tilde{\xi}^{(k)}=\xi^{(k)}$, for $k=1,2$. Therefore, from \ref{identi_Eq_shared_Gamma_frailty_model}, we get 
$$\left[1 + \sigma^2 H^{(k)}_{0} (t_k;\boldsymbol{\xi^{(k)}})\right]^{1 + \frac{1}{\sigma^2}} = \left[1 + \tilde{\sigma}^{2} H^{(k)}_{0}(t_k;\boldsymbol{\xi^{(k)}})\right]^{1 + \frac{1}{\tilde{\sigma}^{2}}},$$
for all $t_{k} > 0$ and $k=1,2$. \\

Note that, for fixed $k$, $H^{(k)}_{0}(t_k;\boldsymbol{\xi^{(k)}})$  is the cumulative hazard function of a continuous random variable. Therefore, there exists a unique value $t^{\ast}_k$ for which $H^{(k)}_{0}(t^{\ast}_k;\boldsymbol{\xi^{(k)}}) = 1$. Taking limit as $t_{k} \to t^{\ast}_{k}$, we get
$$\left[1 + \sigma^2\right]^{1 + \frac{1}{\sigma^2}} = \left[1 + \tilde{\sigma}^{2}\right]^{1 + \frac{1}{\tilde{\sigma}^{2}}}.$$
It can  be checked that the function $g(x) = (1+x)^{1+\frac{1}{x}}$, for $x>0$, is a monotonically increasing function with $\lim_{x \to 0+}g(x)=e$ and hence is a one-to-one function. Therefore, the above equality implies $\sigma=\tilde{\sigma}$. Hence the shared Gamma frailty model (\ref{shared_Gamma_frailty_model}) is identifiable.  

\end{proof}

Let us now investigate the cross-ratio function $CR_{j_{1}j_{2}}(t_1,t_2;\boldsymbol{\xi},\sigma)$ for the shared Gamma frailty model. Firstly, 
\begin{align*}
   \int\limits_{t_2}^{\infty}\sum\limits_{j_{2} = 1}^{L_2}f_{j_{1}j_{2}}(t_1,u_2;\boldsymbol{\xi},\sigma^2)du_2 &=  \int\limits_{t_2}^{\infty}\sum\limits_{j_{2} = 1}^{L_2}\frac{(1 + \sigma^2)h^{(1)}_{0j_{1}}(t_1;\boldsymbol{\xi^{(1)}})h^{(2)}_{0j_{2}}(u_2;\boldsymbol{\xi^{(2)})}}{\left[1 + \sigma^2\left(H^{(1)}_{0}(t_1;\boldsymbol{\xi^{(1)}}) + H^{(2)}_{0}(u_2;\boldsymbol{\xi^{(2)}})\right)\right]^{2 + \frac{1}{\sigma^2}}}du_2\\
   &= (1 + \sigma^2)h^{(1)}_{0j_{1}}(t_1;\boldsymbol{\xi^{(1)}})\int\limits_{t_2}^{\infty}\frac{h^{(2)}_{0}(u_2;\boldsymbol{\xi^{(2)}})}{\left[1 + \sigma^2\left(H^{(1)}_{0}(t_1;\boldsymbol{\xi^{(1)}}) + H^{(2)}_{0}(u_2;\boldsymbol{\xi^{(2)}})\right)\right]^{2 + \frac{1}{\sigma^2}}}du_2\\
   &= \frac{h^{(1)}_{0j_1}(t_1;\boldsymbol{\xi^{(1)}})}{\left[1 + \sigma^2 \sum_{k=1}^2 
   H^{(k)}_{0}(t_k;\boldsymbol{\xi^{(k)}})\right]^{1 + \frac{1}{\sigma^2}}},
\end{align*}
using the transformation $z=1 + \sigma^2\left(H^{(1)}_{0}(t_1;\boldsymbol{\xi^{(1)}}) + H^{(2)}_{0}(u_2;\boldsymbol{\xi^{(2)}})\right)$, where $h^{(2)}_{0}(u_2;\boldsymbol{\xi^{(2)}})= \sum\limits_{j = 1}^{L_2} h^{(2)}_{0j}(u_2;\boldsymbol{\xi^{(2)}})$, for $j_1 = 1,2,\cdots,L_1$ and $t_{1} > 0$. 
Similarly, we have 
\begin{center} $\displaystyle{\int\limits_{t_1}^{\infty}\sum\limits_{j_{1} = 1}^{L_1}f_{j_{1}j_{2}}(u,t_2;\boldsymbol{\xi},\sigma^2)du = \frac{h^{(2)}_{0j_2}(t_2;\boldsymbol{\xi^{(2)}})}{\left[1 + \sigma^2 \sum_{k=1}^2 
   H^{(k)}_{0}(t_k;\boldsymbol{\xi^{(k)}})\right]^{1 + \frac{1}{\sigma^2}}},}$
\end{center}
for $j_2 = 1,2,\cdots,L_2$ and $t_{2} > 0$. Then, we have the expression for $CR_{j_{1}j_{2}}(t_1,t_2;\boldsymbol{\xi},\sigma)$ as 
$$\frac{\frac{(1 + \sigma^2)\prod_{k=1}^2 
h^{(k)}_{0j_k}(t_k;\boldsymbol{\xi^{(k)}})}{\left[1 + \sigma^2 \sum_{k=1}^2 
   H^{(k)}_{0}(t_k;\boldsymbol{\xi^{(k)}})\right]^{2 + \frac{1}{\sigma^2}}} \times \left[1 + \sigma^2 \sum_{k=1}^2 
   H^{(k)}_{0}(t_k;\boldsymbol{\xi^{(k)}})\right]^
   {- \frac{1}{\sigma^2}}}{\frac{h^{(1)}_{0j_1}(t_1;\boldsymbol{\xi^{(1)}})}{\left[1 + \sigma^2 \sum_{k=1}^2 
   H^{(k)}_{0}(t_k;\boldsymbol{\xi^{(k)}})\right]^
   {1 + \frac{1}{\sigma^2}}} \times \frac{h^{(2)}_{0j_2}(t_2;\boldsymbol{\xi^{(2)}})}{\left[1 + \sigma^2 \sum_{k=1}^2 
   H^{(k)}_{0}(t_k;\boldsymbol{\xi^{(k)}})\right]^
   {1 + \frac{1}{\sigma^2}}}} \\
    = 1 + \sigma^2,$$
for $t_k>0,\ j_{k} = 1,\cdots,L_{k}$ and $k = 1,2$. 
Therefore, the failure at time $t_1$ due to cause $j_1$ for the first 
individual and the failure at time $t_2$ due to cause $j_2$ for the second individual are positively associated and this association measure is constant for any two times $t_1,t_2$ and two causes $j_1,j_2$ for the first and the second individual,  respectively. Recall that $\sigma^2$ is the variance of the shared Gamma frailty variable $\epsilon$. Hence, larger the variance of shared frailty, higher is the value of the cross-ratio function.

\subsection{Correlated Gamma frailty model}  

In the correlated Gamma frailty model, there are two different, but correlated, frailty variables $\epsilon^{(k)},\ k=1,2$, for the two individuals in a pair incorporating individual effects. Since the frailty variables are not specific to competing risks, this model also allows the possibility of two entirely different sets of competing risks for the two individuals. The correlated Gamma frailty model can also be applied for the example of cancer incidence times of father and son pair, discussed in the previous sub-section, capturing the possible dependence and unobserved heterogeneity between the father and son through the pair of frailty variables $(\epsilon^{(1)},\epsilon^{(2)})$.\\

The correlated Gamma frailty model is given by  
\begin{equation}\label{correlated Gamma frailty}
\lambda^{(k)}_{j}\big(t_k;\boldsymbol{\xi}| \epsilon^{(k)}\big) = h^{(k)}_{0j}\big(t_k;\boldsymbol{\xi}\big)\epsilon^{(k)}, 
\end{equation}
for $t_{k} > 0, j = 1,\cdots,L_{k}$ and $k = 1,2$. The two correlated frailty variables $\epsilon^{(k)},\ k=1,2$, are 
constructed in such a way that they have a common part accounting for the correlation between them and the other parts accounting for the individual effects (See Yashin et al., $1995$). We write $\epsilon^{(k)} = \frac{\mu_0}{\mu_k}Y_{0} + Y_{k}$, for $k = 1,2$, where $Y_i$ follows a $\text{Gamma}(\mu_i,\kappa_i)$ distribution, for $i=0,1,2$. We also assume that the $Y_i$'s are independent random variables 
with $\mu_0,\mu_1,\mu_2,\kappa_0,\kappa_1,\kappa_2 > 0$ satisfying $\mu_{k} = \kappa_{0} + \kappa_{k}$ for $k = 1,2$. These two relationships between the parameters ensure that the frailty means are equal to unity. The association between $\epsilon^{(1)}$ and $\epsilon^{(2)}$ is modeled through the common random variable $Y_{0}$. Note that there are three independent parameters in this model. We consider the re-parametrization of these parameters given by 
$$\sigma_1=\frac{1}{\sqrt{\kappa_{0} + \kappa_{1}}},\ \sigma_2=\frac{1}{\sqrt{\kappa_{0} + \kappa_{2}}},\ \rho=\frac{\kappa_0}{\sqrt{\kappa_{0} + \kappa_{1}}\sqrt{\kappa_{0} + \kappa_{2}}},$$ 
such that $\sigma_k$ is the standard deviation of $\epsilon^{(k)}$ and $\rho$ is the correlation coefficient between $\epsilon^{(1)}$ and $\epsilon^{(2)}$ satisfying the condition $\rho < \min\big(\frac{\sigma_1}{\sigma_2},\frac{\sigma_2}{\sigma_1}\big)$. The frailty parameter space is, therefore, given by
$$\boldsymbol{\Theta} = \{\boldsymbol{\theta}=(\sigma_1,\sigma_2,\rho): \sigma_{1} > 0, \sigma_{2} > 0, 0 < \rho < \min\big(\frac{\sigma_1}{\sigma_2},\frac{\sigma_2}{\sigma_1}\big)\}.$$
The joint density $g(\epsilon^{(1)},\epsilon^{(2)};\boldsymbol{\theta})$ of $(\epsilon^{(1)},\epsilon^{(2)})$ can be written in terms of the three independent Gamma densities of $Y_0,\ Y_1$ and $Y_2$. \\

As before, given the frailty variable $\epsilon^{(k)}$, we have 
$$S^{(k)}(t_{k};\boldsymbol{\xi}|\epsilon^{(k)}) = \exp{\bigg[-\epsilon^{(k)}H^{(k)}_{0}(t_{k};\boldsymbol{\xi})\bigg]} = \exp{\bigg[-\Big(\frac{\mu_0}{\mu_k}Y_{0} + Y_{k}\Big)H^{(k)}_{0}(t_{k};\boldsymbol{\xi})\bigg]}$$
and 
$$F^{(k)}_{j}(t_k;\boldsymbol{\xi} | \epsilon^{(k)}) = \int\limits_{0}^{t_k}h^{(k)}_{0j}(u_k;\boldsymbol{\xi})\epsilon^{(k)}\exp{\bigg[-\epsilon^{(k)}H^{(k)}_{0}(u_k;\boldsymbol{\xi})\bigg]}du_k,$$
for $t_k>0,\ j = 1,\cdots,L_{k}$ and $k = 1,2$. Then, the unconditional joint sub-distribution function $F_{j_{1}j_{2}}(t_1,t_2;\boldsymbol{\xi},\boldsymbol{\theta})$ for the correlated Gamma frailty model is obtained as, using conditional independence and integrating with respect to the densities of $Y_0,\ Y_1$ and $Y_2$, 
\begin{align*}
 &
\int\limits_{0}^{t_1}\int\limits_{0}^{t_2}
\frac{\prod_{k=1}^2 \sigma^{2}_{k}h^{(k)}_{0j_k}(u_k;\boldsymbol{\xi})}{\Big(A(u_1,u_2;\boldsymbol{\xi},\sigma_1,\sigma_2)\Big)^{\frac{\rho}{\sigma_1\sigma_2}} \prod_{k=1}^2 \Big(B_k(u_k;\boldsymbol{\xi},\sigma_k)\Big)^{\left(\frac{1}{\sigma_k^2}-\frac{\rho}{\sigma_1\sigma_2}\right)}} \times 
\Bigg[\frac{\frac{\rho}{\sigma_1\sigma_2}(\frac{\rho}{\sigma_1\sigma_2} + 1)}{\Big(A(u_1,u_2;\boldsymbol{\xi},\sigma_1,\sigma_2)\Big)^{2}} + \\
& \qquad\qquad \sum_{k=1}^2 \frac{\frac{\rho}{\sigma_1\sigma_2}\left(\frac{1}{\sigma_k^2} - \frac{\rho}{\sigma_1\sigma_2}\right)} 
{A(u_1,u_2;\boldsymbol{\xi},\sigma_1,\sigma_2)
B_k(u_k;\boldsymbol{\xi},\sigma_k)} + \prod_{k=1}^2 
\frac{\left(\frac{1}{\sigma_k^2} - \frac{\rho}{\sigma_1\sigma_2}\right)}
{B_k(u_k;\boldsymbol{\xi},\sigma_k)}\Bigg]du_2du_1,
\end{align*}
where $A(u_1,u_2;\boldsymbol{\xi},\sigma_1,\sigma_2) = 1+\sum_{k=1}^2\sigma^{2}_{k}H^{(k)}_{0}(u_k;\boldsymbol{\xi})$, $B_k(u_k;\boldsymbol{\xi},\sigma_k) = 1+ \sigma^{2}_{k}H^{(k)}_{0}(u_k;\boldsymbol{\xi})$ for $u_k> 0$ and $k=1,2$.\\

The unconditional joint survival function is given by, using conditional independence and integrating with respect to the densities of $Y_0,\ Y_1$ and $Y_2$ as before,
$$S(t_1,t_2;\boldsymbol{\xi},\boldsymbol{\theta}) = \left(A(t_1,t_2;\boldsymbol{\xi},\sigma_1,\sigma_2)\right)^{-\frac{\rho}{\sigma_{1}\sigma_{2}}}
\prod_{k=1}^2 \Big(B_k(t_k;\boldsymbol{\xi},\sigma_k)\Big)^{-\left(\frac{1}{\sigma_k^2}-\frac{\rho}{\sigma_1\sigma_2}\right)},$$
for all $t_1,t_2 > 0$. 
The unconditional joint sub-density function $f_{j_1,j_2}(t_1,t_2;\boldsymbol{\xi},\boldsymbol{\theta})$ is given by 
\begin{align*}
& \prod_{k=1}^2 \left(\sigma^{2}_{k}h^{(k)}_{0j_k}(t_k;\boldsymbol{\xi})\right)
S(t_1,t_2;\boldsymbol{\xi},\sigma_1,\sigma_2)
\Bigg[\frac{\frac{\rho}{\sigma_1\sigma_2}\left(\frac{\rho}{\sigma_1\sigma_2} + 1\right)}{\Big(A(t_1,t_2;\boldsymbol{\xi},\sigma_1,\sigma_2)\Big)^{2}} + \\
&\qquad \sum_{k=1}^2 \frac{\frac{\rho}{\sigma_1\sigma_2}\left(\frac{1}{\sigma_k^2} - \frac{\rho}{\sigma_1\sigma_2}\right)} 
{A(t_1,t_2;\boldsymbol{\xi},\sigma_1,\sigma_2)
B_k(t_k;\boldsymbol{\xi},\sigma_k)} + \prod_{k=1}^2 
\frac{\left(\frac{1}{\sigma_k^2} - \frac{\rho}{\sigma_1\sigma_2}\right)}
{B_k(t_k;\boldsymbol{\xi},\sigma_k)}\Bigg], 
\end{align*}
for all $t_{k},\ j_{k} = 1,\cdots,L_{k}$ and $k = 1,2$.\\ 

\begin{definition}
    The correlated Gamma frailty model $(\ref{correlated Gamma frailty})$ is identifiable within families $\boldsymbol{\Xi}\times\boldsymbol{\Theta}$ with 
$\boldsymbol{\Theta}$ as described above in this sub-section if, for some $\boldsymbol{\xi},\boldsymbol{\tilde{\xi}} \in \boldsymbol{\Xi}$ and $(\sigma_1,\sigma_2,\rho),(\tilde{\sigma_1},\tilde{\sigma_2},\tilde{\rho}) \in \boldsymbol{\Theta}$, the equality of 
$F_{j_{1}j_{2}}(t_1,t_2;\boldsymbol{\xi},\sigma_1,\sigma_2,\rho)$ and $F_{j_{1}j_{2}}(t_1,t_2;\boldsymbol{\tilde{\xi}},\tilde{\sigma_1},\tilde{\sigma_2},\tilde{\rho})$ with the above expressions, for all $t_{k} > 0,\ j_{k} = 1,\cdots,L_{k}$ and $k = 1,2$, implies $\boldsymbol{\xi} =\tilde{\boldsymbol{\xi}}$ and $\sigma_{k} =\tilde{\sigma}_{k}$, for $k = 1,2$ and $\rho = \tilde{\rho}$.
\end{definition}

For $k=1,2$, integrating $F^{(k)}_{j}(t_k;\boldsymbol{\xi} | \epsilon^{(k)})$ with respect to the distribution of $\epsilon^{(k)}$, or $Y_0$ and $Y_k$, we get the corresponding unconditional sub-distribution function $F_j^{(k)}(t_k;\boldsymbol{\xi},\sigma_1,\sigma_2,\rho)$ as 
\begin{align*}
&\int\limits_{0}^{\infty}\int\limits_{0}^{\infty}\bigg(\int\limits_{0}^{t_k}h^{(k)}_{0j}(u_k;\boldsymbol{\xi})\Big(\frac{\mu_0}{\mu_k}y_{0} + y_{k}\Big)\exp{\bigg[-\Big(\frac{\mu_0}{\mu_k}y_{0} + y_{k}\Big)H^{(k)}_{0}(u_k;\boldsymbol{\xi})\bigg]}du_k\bigg)\\
    &\qquad\qquad\times \frac{\mu^{\kappa_0}_0}{\Gamma(\kappa_0)}\frac{\mu^{\kappa_k}_k}{\Gamma(\kappa_k)}\exp{[-\mu_{0}y_{0}-\mu_{k}y_{k}]}y^{\kappa_{0} - 1}_{0}y^{\kappa_{k} - 1}_{k}dy_{0}dy_{k}\\
    &= \int\limits_{0}^{t_k}h^{(k)}_{0j}(u_k;\boldsymbol{\xi})[I_{1j}(u_k) + I_{2j}(u_k)]du_k, 
\end{align*}
say, where
\begin{align*}
   I_{1j}(u_k) &= \frac{\mu_0}{\mu_k} \frac{\mu^{\kappa_0}_0}{\Gamma(\kappa_0)}\frac{\mu^{\kappa_k}_k}{\Gamma(\kappa_k)}\int\limits_{0}^{\infty}\int\limits_{0}^{\infty}\exp{\bigg[-\mu_{0}y_{0}\Big(\frac{H^{(k)}_{0}(u_k;\boldsymbol{\xi})}{\mu_k} + 1\Big)-\mu_{k}y_{k}\Big(\frac{H^{(k)}_{0}(u_k;\boldsymbol{\xi})}{\mu_k} + 1\Big)\bigg]}y^{\kappa_{0} + 1 - 1}_{0}y^{\kappa_{k} - 1}_{k}dy_{0}dy_{k}\\ 
   &= \frac{\kappa_0}{\mu_k} \Big[\frac{H^{(k)}_{0}(u_k;\boldsymbol{\xi})}{\mu_k} + 1\Big]^{-\kappa_0-\kappa_{k}-1} = \frac{\kappa_0}{\mu_k} \Big[\frac{H^{(k)}_{0}(u_k;\boldsymbol{\xi})}{\mu_k} + 1\Big]^{-\mu_{k}-1}
\end{align*}
and
\begin{align*}
    I_{2j}(u_k) &= \frac{\mu^{\kappa_0}_0}{\Gamma(\kappa_0)}\frac{\mu^{\kappa_k}_k}{\Gamma(\kappa_k)}\int\limits_{0}^{\infty}\int\limits_{0}^{\infty}\exp{\bigg[-\mu_{0}y_{0}\Big(\frac{H^{(k)}_{0}(u_k;\boldsymbol{\xi})}{\mu_k} + 1\Big)-\mu_{k}y_{k}\Big(\frac{H^{(k)}_{0}(u_k;\boldsymbol{\xi})}{\mu_k} + 1\Big)\bigg]}y^{\kappa_{0}- 1}_{0}y^{\kappa_{k} + 1 - 1}_{k}dy_{0}dy_{k}\\
    &= \frac{\kappa_k}{\mu_k}\Big[\frac{H^{(k)}_{0}(u_k;\boldsymbol{\xi})}{\mu_k} + 1\Big]^{-\kappa_{0}-\kappa_{k}-1} = \frac{\kappa_k}{\mu_k}\Big[\frac{H^{(k)}_{0}(u_k;\boldsymbol{\xi})}{\mu_k} + 1\Big]^{-\mu_{k}-1}.
\end{align*}
Note that this unconditional sub-distribution function depends only on $\xi^{(k)}$ and $\sigma_{k}$ and is given by 
\begin{align*}
  F^{(k)}_{j}(t_k;\boldsymbol{\xi^{(k)}},\sigma_{k})  &= \int\limits_{0}^{t_k}h^{(k)}_{0j}(u_k;\boldsymbol{\xi})\Bigg(\frac{\kappa_{0} + \kappa_{k}}{\mu_k} \Big[\frac{H^{(k)}_{0}(u_k;\boldsymbol{\xi})}{\mu_k} + 1\Big]^{-\mu_{k}-1}\Bigg)du_k\\
 &= \int\limits_{0}^{t_k}\frac{a(\gamma^{(k)}_{j},\alpha^{(k)}_{j})u_k^{\gamma^{(k)}_{j} - 1}b(u_k;\gamma^{(k)}_{j},\alpha^{(k)}_{j})du_k}{\big[1 + \sigma^{2}_{k} H^{(k)}_{0}(u_k;\boldsymbol{\xi^{(k)}})\big]^{1 + \frac{1}{\sigma^{2}_{k}}}}
\end{align*}
for $t_{k} > 0,\ j = 1,\cdots,L_{k}$ and $k = 1,2$. This expression is similar to \ref{marg_subdis_shared_Gamma_frailty}, the unconditional sub-distribution function for the shared Gamma frailty model (\ref{shared_Gamma_frailty_model}) with $\sigma$ replaced by $\sigma_k$. \\

\begin{theorem}
The correlated Gamma frailty model (\ref{correlated Gamma frailty}), with the cause-specific baseline hazard functions belonging to the family defined in (\ref{parametric_family_haz}), is identifiable within $\boldsymbol{\Xi}\times\boldsymbol{\Theta}$.
\end{theorem}

\begin{proof}
According to \textbf{Definition} $\boldsymbol{3}$, the equality of unconditional joint sub-distribution functions also implies the equality of unconditional marginal sub-distribution functions evaluated at $(\boldsymbol{\xi},\boldsymbol{\theta})$ and $(\boldsymbol{\tilde{\xi}},\boldsymbol{\tilde{\theta}})$. Since the unconditional marginal sub-distribution functions have the same form as those of the shared Gamma frailty model, following the same technique as used to prove \textbf{Theorem} $\boldsymbol{2.1}$, we have
$$\xi^{(k)}_{j} = \tilde{\xi}^{(k)}_{j}\quad\mbox{ and }\quad \sigma_{k} = \tilde{\sigma}_{k},$$
for $j=1,\cdots,L_k,\ k = 1,2$. \\

Next, using the equality of unconditional joint survival functions $S(t_1,t_2;\boldsymbol{\xi},\sigma_1,\sigma_2,\rho)$ and $S(t_1,t_2;\boldsymbol{\tilde{\xi}},\tilde{\sigma}_1,\tilde{\sigma}_2,\tilde{\rho})$, along with $\xi^{(k)}_{j} = \tilde{\xi}^{(k)}_{j}$ and $\sigma_{k} = \tilde{\sigma}_{k}$ for $j = 1,\cdots,L_{k},\  k = 1,2$, we have 
\begin{multline*}
    \big[1 + \sum_{k=1}^2 
    \sigma^{2}_{k} H^{(k)}_{0}(t_k;\boldsymbol{\xi}) \big]^{-\frac{\rho}{\sigma_{1}\sigma_{2}}}
    \prod_{k=1}^2\big[1 +     
    \sigma^{2}_{k} H^{(k)}_{0}(t_k;\boldsymbol{\xi})\big]^{\big(\frac{\rho}{\sigma_{1}\sigma_{2}} - \frac{1}{\sigma^{2}_k}\big)}
    \\
= \big[1 + \sum_{k=1}^2 
    \sigma^{2}_{k} H^{(k)}_{0}(t_k;\boldsymbol{\xi}) \big]^{-\frac{\tilde{\rho}}{\sigma_{1}\sigma_{2}}}
    \prod_{k=1}^2\big[1 +     
    \sigma^{2}_{k} H^{(k)}_{0}(t_k;\boldsymbol{\xi})\big]^{\big(\frac{\tilde{\rho}}{\sigma_{1}\sigma_{2}} - \frac{1}{\sigma^{2}_k}\big)},
\end{multline*}
for all $t_{1},t_{2} > 0$. This implies 
\begin{equation*}
    \bigg[\frac{\prod_{k=1}^{2}\left(1 + \sigma^{2}_{k} H^{(k)}_{0}(t_k;\boldsymbol{\xi})\right)}{1 + \sum_{k=1}^2 
    \sigma^{2}_{k} H^{(k)}_{0}(t_k;\boldsymbol{\xi})}
    \bigg]^{\frac{\rho - \tilde{\rho}}{\sigma_{1}\sigma_{2}}} = 1,
\end{equation*}
for all $t_{1},t_{2} > 0$. Hence, we get $\rho = \tilde{\rho}$. Therefore, the correlated Gamma frailty model (\ref{correlated Gamma frailty}) is identifiable. 
\end{proof}

In this case, the cross-ratio function $CR_{j_{1}j_{2}}(t_1,t_2;\boldsymbol{\xi},\sigma_{1},\sigma_{2},\rho)$ 
between the occurrence of failure at time $t_1$ due to cause $j_1$ for first individual and the occurrence of failure at time $t_2$ due to cause $j_2$ for the second individual is given by 
\begin{align*}
& \left[\frac{\left(\prod_{k=1}^2 
\sigma^{2}_{k}h^{(k)}_{0j_k}(t_k;\boldsymbol{\xi})\right) 
\left(S(t_1,t_2;\boldsymbol{\xi},\sigma_{1},\sigma_{2},\rho)\right)^{2}}
{\left(\int\limits_{t_2}^{\infty}\sum\limits_{j = 1}^{L_2}f_{j_1j}(t_1,u_2;\boldsymbol{\xi},\sigma_{1},\sigma_{2},\rho)du_2\right) \times \left(\int\limits_{t_1}^{\infty}\sum\limits_{j = 1}^{L_1}f_{jj_2}(u_1,t_2;\boldsymbol{\xi},\sigma_{1},\sigma_{2},\rho)du_1\right)}\right]\times \\
&\qquad \Bigg[\frac{\frac{\rho}{\sigma_1\sigma_2}\left(\frac{\rho}{\sigma_1\sigma_2} + 1\right)}{\big(A(t_1,t_2;\boldsymbol{\xi},\sigma_{1},\sigma_{2})\big)^{2}} + \sum_{k=1}^2 
\frac{\frac{\rho}{\sigma_1\sigma_2}\left(\frac{1}{\sigma^{2}_{k}} - \frac{\rho}{\sigma_1\sigma_2}\right)}{A(t_1,t_2;\boldsymbol{\xi},\sigma_{1},\sigma_{2})B_k(t_k;\boldsymbol{\xi},\sigma_{k})} + \prod_{k=1}^2
 \frac{\frac{1}{\sigma^{2}_{k}} - \frac{\rho}{\sigma_1\sigma_2}}{B_k(t_k;\boldsymbol{\xi},\sigma_{k})}\Bigg]
\end{align*}
for $t_{k} > 0,\ j_{k} = 1,\cdots,L_{k}$ and $k = 1,2$.\\

\begin{figure}[h]
    \centering
\includegraphics[scale=0.8]{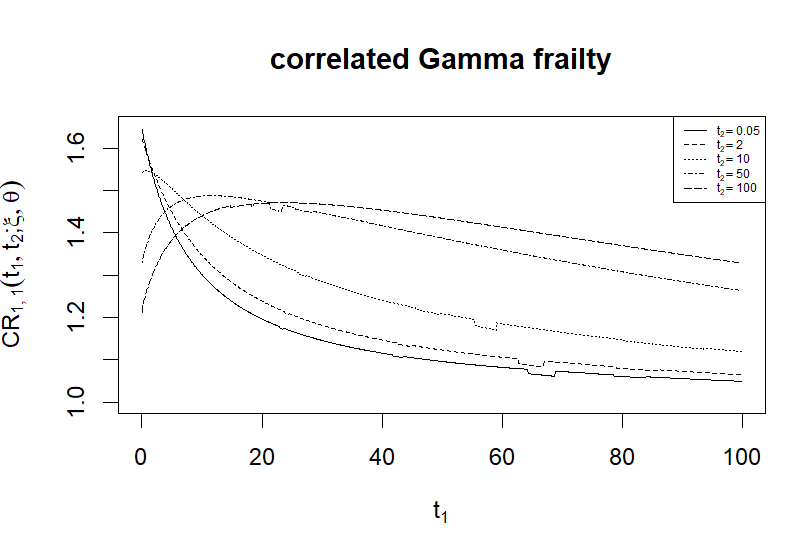}
    \caption{The cross ratio function corresponding to correlated Gamma frailty}
    \label{fig:Cr-ratio-correlated-Gamma}
\end{figure}

In case of constant cause-specific baseline hazard (that is, $h^{(k)}_{0j}(t_{k};\boldsymbol{\xi}) = \alpha^{(k)}_{j}$, for $j = 1,\cdots,L_{k}$ and $k = 1,2$), the above expression for the cross-ratio function simplifies to  

\begin{align*}
    & \frac{1}{\prod_{k=1}^2 \sigma^{2}_{k}\alpha^{(k)}}\times \frac{\left(S(t_1,t_2;\boldsymbol{\xi},\boldsymbol{\theta})^2\right)D(t_1,t_2;\boldsymbol{\xi},\boldsymbol{\theta})}{\int\limits_{t_1}^{\infty}S(u_1,t_2;\boldsymbol{\xi},\boldsymbol{\theta})D(u_1,t_2;\boldsymbol{\xi},\boldsymbol{\theta})du_1 \times \int\limits_{t_2}^{\infty}S(t_1,u_2;\boldsymbol{\xi},\boldsymbol{\theta})D(t_1,u_2;\boldsymbol{\xi},\boldsymbol{\theta})du_2}
\end{align*}
where 
\begin{align*}
    D(t_1,t_2;\boldsymbol{\xi},\boldsymbol{\theta}) &= 
    \Bigg[\frac{\frac{\rho}{\sigma_{1}\sigma_{2}}(\frac{\rho}{\sigma_{1}\sigma_{2}} + 1)}{\left(A(t_1,t_2;\boldsymbol{\xi},\sigma_1,\sigma_2)\right)^2} + \sum_{k=1}^2
    \frac{\frac{\rho}{\sigma_{1}\sigma_{2}}(\frac{1}{\sigma^{2}_k} - \frac{\rho}{\sigma_{1}\sigma_{2}})}{A(t_1,t_2;\boldsymbol{\xi},\sigma_1,\sigma_2)B_k(t_k;\boldsymbol{\xi},\sigma_k)} + \prod_{k=1}^2 
 \frac{\frac{1}{\sigma^{2}_{k}} - \frac{\rho}{\sigma_1\sigma_2}}{B_k(t_k;\boldsymbol{\xi},\sigma_{k})}
 \Bigg].
\end{align*}
It is difficult to find this cross-ratio function analytically, even to obtain a suitable lower bound. Note that this cross-ratio function is the same for all $(j_1,j_2)$. We numerically evaluate this for constant cause-specific baseline hazards and 
different values of $(t_1,t_2)$,as presented in Figure \ref{fig:Cr-ratio-correlated-Gamma}. We assume two competing risks for each individual with $\boldsymbol{\xi^{(1)}} = (\alpha^{(1)}_{1},\alpha^{(1)}_{2}) = (0.2,0.25)$ and $\boldsymbol{\xi^{(2)}} = (\alpha^{(2)}_{1},\alpha^{(2)}_{2}) = (0.15,0.1)$. 
The frailty parameter vector $\boldsymbol{\theta}=(\sigma_1,\sigma_2,\rho)$ is taken as $(0.95,0.85,0.8)$. In figure \ref{fig:Cr-ratio-correlated-Gamma}, we give  plots of only the cross-ratio function $CR_{1,1}(t_1,t_2;\boldsymbol{\xi},\boldsymbol{\theta})$ against $t_1$ for five different values of $t_2$ as $(0.05,0.2,0.5,0.9,2)$. 

For this particular set of model parameters, the Figure \ref{fig:Cr-ratio-correlated-Gamma} shows that the cross-ratio function is always greater than unity. We have computed the cross-ratio function for other parameter values and observed the same (not reported here). Therefore, failure times of the two individuals due to some fixed causes seem to be positively correlated. 
\newpage
\subsection{Shared cause-specific Gamma frailty model}

Contrary to shared and correlated Gamma frailty models  (\ref{shared_Gamma_frailty_model}) and (\ref{correlated Gamma frailty}), discussed in the previous two sub-sections, the shared cause-specific frailty model involves a common frailty variable for the two individuals in a pair, depending on the cause of failure as well. As a result, the set of competing risks for the two individuals needs to be the same, so that the common frailty variable for the cause $j$ may be denoted by $\epsilon_{j}$, for $j=1,\cdots,L$ with $L_1=L_2=L$. In the example of cancer incidence times of father and son pair, the incidence times may be differently associated for the different cancer sites, but the dependence for a particular site, say $j$th, is modeled through a common shared frailty $\epsilon_{j}$. \\ 

The shared cause-specific Gamma frailty model is given by 
\begin{equation}\label{shared cause specific Gamma frailty} 
\lambda^{(k)}_{j}\big(t_k;\boldsymbol{\xi}| \epsilon_j\big) = h^{(k)}_{0j}\big(t_k;\boldsymbol{\xi}\big)\epsilon_{j} 
\end{equation}
for all $t_{k} > 0,\ j = 1,\cdots,L$ and $k = 1,2$. We write $\boldsymbol{\epsilon}=(\epsilon_1,\cdots,\epsilon_L)$ and assume that the $\epsilon_j$'s independently follow $\text{Gamma}(\frac{1}{\sigma^{2}_j},\frac{1}{\sigma^{2}_j})$ distribution with $\sigma_{j} > 0$, so that $\mathbb{E}(\epsilon_j)$ is unity, for $j = 1,\cdots,L$. The frailty parameter space  is, therefore, given by
$$\boldsymbol{\Theta} = \{\boldsymbol{\theta} = (\sigma_1,\cdots,\sigma_L):\sigma_{j} > 0,j = 1,2,\cdots,L\}.$$

As before, conditional on $\boldsymbol{\epsilon}$, the survival function of $k$th individual is 
$$S^{(k)}(t_k;\boldsymbol{\xi}|\boldsymbol{\epsilon}) = \exp{\Bigg[-\sum\limits_{j=1}^{L}H^{(k)}_{0j}(t_k;\boldsymbol{\xi})\epsilon_{j}\Bigg]}$$
and the $j$th sub-distribution function of $k$th individual is
$$F^{(k)}_{j}(t_k;\boldsymbol{\xi}|\boldsymbol{\epsilon}) = \int\limits_{0}^{t_k}h^{(k)}_{0j}(u_k;\boldsymbol{\xi})\epsilon_{j}\exp{\Bigg[-\sum\limits_{j'=1}^{L}H^{(k)}_{0j'}(u_k;\boldsymbol{\xi})\epsilon_{j'}\Bigg]}du_k,$$
for all $t_{k} > 0,\ j = 1,2,\cdots,L$ and $k = 1,2$. The unconditional joint sub-distribution function, $F_{j_{1}j_{2}}(t_1,t_2;\boldsymbol{\xi},\boldsymbol{\theta})$, is given by 
$$\int\limits_{0}^{\infty}\cdots\int\limits_{0}^{\infty}\Bigg[\int\limits_{0}^{t_1}\int\limits_{0}^{t_2} \left(\prod_{k=1}^2 
h^{(k)}_{0j_k}(u_k;\boldsymbol{\xi})\epsilon_{j_k}\right)
\times\exp{\Bigg(-\sum\limits_{j=1}^{L} \epsilon_{j} \sum_{k=1}^2 
H^{(k)}_{0j}(u_k;\boldsymbol{\xi}) 
\Bigg)}du_2du_1\Bigg]g(\boldsymbol{\epsilon};\boldsymbol{\theta})d\boldsymbol{\epsilon},$$ 
for all $t_{k} > 0,\ j_{k} = 1,\cdots,L$ and $k = 1,2$, where $g(\boldsymbol{\epsilon};\boldsymbol{\theta})$ is the joint density of $L$ independent Gamma variables $\epsilon_1,\cdots,\epsilon_L$. 
The expression of $F_{j_{1}j_{2}}(t_1,t_2;\boldsymbol{\xi},\boldsymbol{\theta})$ simplifies to 
    \begin{align*}
      F_{jj}(t_1,t_2;\boldsymbol{\xi},\boldsymbol{\theta}) &= (1 + \sigma^{2}_j)\int\limits_{0}^{t_1}\int\limits_{0}^{t_2} \left(\prod_{k=1}^2 h^{(k)}_{0j}(u_k;\boldsymbol{\xi})\right)\times 
      \bigg[1 + \sigma^{2}_{j} \sum_{k=1}^2 H^{(k)}_{0j}(u_k;\boldsymbol{\xi})\bigg]^{-2} \\
       &\qquad\times 
      \prod\limits_{j'=1}^{L} \bigg[1 + \sigma^{2}_{j'} \sum_{k=1}^2 H^{(k)}_{0j'}(u_k;\boldsymbol{\xi})\bigg]^{-\frac{1}{\sigma^{2}_{j'}}}du_2du_1,
    \end{align*}
for $j_1=j_2=j$ (say), and 
    \begin{align*}
    F_{j_{1}j_{2}}(t_1,t_2;\boldsymbol{\xi},\boldsymbol{\theta}) &= \int\limits_{0}^{t_1}\int\limits_{0}^{t_2} \prod_{k=1}^2 \left(
    h^{(k)}_{0j_k}(u_k;\boldsymbol{\xi})\right)\times 
    \prod\limits_{k=1}^{2}\bigg[1 + \sigma^{2}_{j_k}\big(H^{(1)}_{0j_k}(u_1;\boldsymbol{\xi}) + H^{(2)}_{0j_k}(u_2;\boldsymbol{\xi})\big)\bigg]^{-1} \\ 
     &\qquad\times 
    \prod\limits_{j'=1}^{L} \bigg[1 + \sigma^{2}_{j'}\sum_{k=1}^2 
    H^{(k)}_{0j'}(u_k;\boldsymbol{\xi})\bigg]^{-\frac{1}{\sigma^{2}_{j'}}}du_2du_1, 
    \end{align*}
   for $j_{1} \neq j_{2}$ and $t_{k} > 0,\ j_{k} = 1,2,\cdots,L,\ k = 1,2$.   Then, the joint sub-density function $f_{j_{1}j_{2}}(t_1,t_2;\boldsymbol{\xi},\boldsymbol{\theta})$ is given by 
    \begin{align*}
      f_{jj}(t_1,t_2;\boldsymbol{\xi},\boldsymbol{\theta}) &= (1 + \sigma^{2}_j) \left(\prod_{k=1}^2 h^{(k)}_{0j}(t_k;\boldsymbol{\xi})\right)\times \bigg[1 + \sigma^{2}_{j}
    \sum_{k=1}^2 H^{(k)}_{0j}(t_k;\boldsymbol{\xi})\bigg]^{-2} \\
       &\qquad\times 
      \prod\limits_{j'=1}^{L} \bigg[1 + \sigma^{2}_{j'} \sum_{k=1}^2 H^{(k)}_{0j'}(t_k;\boldsymbol{\xi})\bigg]^{-\frac{1}{\sigma^{2}_{j'}}}, 
    \end{align*}
   for $j_{1} = j_{2} = j$ (say), and 
    \begin{align*}
    f_{j_{1}j_{2}}(t_1,t_2;\boldsymbol{\xi},\boldsymbol{\theta}) &= \left(\prod_{k=1}^2h^{(k)}_{0j_k}(t_k;\boldsymbol{\xi})\right)\times 
    \prod\limits_{k=1}^{2}\bigg[1 + \sigma^{2}_{j_k}\big(H^{(1)}_{0j_k}(t_1;\boldsymbol{\xi}) + H^{(2)}_{0j_k}(t_2;\boldsymbol{\xi})\big)\bigg]^{-1} \\
     &\qquad\times
    \prod\limits_{j'=1}^{L} \bigg[1 + \sigma^{2}_{j'}\sum_{k=1}^2 
    H^{(k)}_{0j'}(t_k;\boldsymbol{\xi})\bigg]^{-\frac{1}{\sigma^{2}_{j'}}}, 
    \end{align*}
for $j_{1} \neq j_{2}$ and $t_{k} > 0,\ j_{k} = 1,\cdots,L,\ k = 1,2$. 
The unconditional joint survival function is given by
\begin{align*}
    S(t_1,t_2;\boldsymbol{\xi},\boldsymbol{\theta}) &= \int\limits_{0}^{\infty}\cdots\int\limits_{0}^{\infty} \exp{\Bigg[-\sum\limits_{j=1}^{L}\epsilon_{j} \sum_{k=1}^2 
    H^{(k)}_{0j}(t_k;\boldsymbol{\xi})\Bigg]}g(\boldsymbol{\epsilon};\boldsymbol{\theta})d\boldsymbol{\epsilon}\\
    &= \prod\limits_{j=1}^{L}\Big[1 + \sigma^{2}_{j}\sum_{k=1}^2 
    H^{(k)}_{0j}(t_k;\boldsymbol{\xi})\Big]^{-\frac{1}{\sigma^{2}_j}},
\end{align*}
for all $t_{1},t_{2} > 0$. Now, it is easy to check that the unconditional survival function of the $k$th  individual, by letting $t_{k'}\rightarrow 0$ for $k'\neq k$, is given by 
$$S^{(k)}(t_k;\boldsymbol{\xi^{(k)}},\boldsymbol{\theta})= 
\prod\limits_{j=1}^{L}\Big[1 + \sigma^{2}_{j}H^{(k)}_{0j}(t_k; 
\gamma_{j}^{(k)},\alpha_{j}^{(k)})\Big]^{-\frac{1}{\sigma^{2}_j}},$$
depending only on $\boldsymbol{\xi^{(k)}}$ and $\boldsymbol{\theta}$. \\

\begin{definition}
    The shared cause-specific Gamma frailty model $(\ref{shared cause specific Gamma frailty})$ is identifiable within families  $\boldsymbol{\Xi}\times\boldsymbol{\Theta}$ with  
$\boldsymbol{\Theta}$ as described above in this sub-section if, for some $\boldsymbol{\xi},\boldsymbol{\tilde{\xi}} \in \boldsymbol{\Xi}$ and $\boldsymbol{\theta}=(\sigma_1,\cdots,\sigma_L), \boldsymbol{\tilde{\theta}}= (\tilde{\sigma_1},\cdots,\tilde{\sigma_L}) \in \boldsymbol{\Theta}$, the equality of 
$F_{j_{1}j_{2}}(t_1,t_2;\boldsymbol{\xi},\sigma_1,\cdots,\sigma_L)$ and $F_{j_{1}j_{2}}(t_1,t_2;\boldsymbol{\tilde{\xi}},\tilde{\sigma_1},\cdots,\tilde{\sigma_L})$  with the above expressions, for all $t_{k} > 0,\ j_{k} = 1,\cdots,L$ and $k = 1,2$, implies $\boldsymbol{\xi} =\tilde{\boldsymbol{\xi}}$ and $\sigma_{k} =\tilde{\sigma}_{k}$, for $k = 1,\cdots,L$.
\end{definition}

Note that, as in Sub-setion 2.1, the unconditional $j$th sub-distribution function $F^{(k)}_{j}(t_k;\boldsymbol{\xi^{(k)}},\sigma_j)$ 
for the $k$th individual depends only on $\boldsymbol{\xi^{(k)}}$ and $\sigma_j$ and is given by the same expression as that in $(\ref{marg_subdis_shared_Gamma_frailty})$, with $\sigma$ replaced by $\sigma_j$, for $j = 1,\cdots,L$ and $k = 1,2$. \\

\begin{theorem}
The shared cause-specific Gamma frailty model $(\ref{shared cause specific Gamma frailty})$, with the cause-specific baseline hazard functions belonging to the family defined in (\ref{parametric_family_haz}), is identifiable within $\boldsymbol{\Xi}\times\boldsymbol{\Theta}$.
\end{theorem}

\begin{proof}
The equality of the joint unconditional sub-distribution functions with $(\boldsymbol{\xi},\boldsymbol{\theta})$ and $(\boldsymbol{\tilde{\xi}},\boldsymbol{\tilde{\theta}})$ implies equality of the corresponding unconditional sub-density functions (See Sub-section 2.1) $f^{(k)}_{j}(t_k;\boldsymbol{\xi^{(k)}},\boldsymbol{\theta})$ and $f^{(k)}_{j}(t_k;\boldsymbol{\tilde{\xi}^{(k)}},\boldsymbol{\tilde{\theta}})$ 
for $t_{k} > 0, j = 1,\cdots,L$ and $k=1,2$. This gives, after some rearrangements as in (\ref{identi_Eq_shared_Gamma_frailty_model}), 
\begin{equation}\label{shared_cause_specific_frailty_Eq1}
1 = t_k^{\tilde{\gamma}_{j}^{(k)} - \gamma_{j}^{(k)}} \frac{ a(\tilde{\gamma}_{j}^{(k)},\tilde{\alpha}_{j}^{(k)}) b(t_k;\tilde{\gamma}_{j}^{(k)}, \tilde{\alpha}_{j}^{(k)}) 
\Big[1 + \sigma^{2}_{j}H^{(k)}_{0j}(t_k,\gamma_{j}^{(k)},\alpha_{j}^{(k)})\Big] 
\prod\limits_{j'=1}^{L}\Big[1 + \sigma^{2}_{j'}H^{(k)}_{0j'}(t_k,\gamma_{j'}^{(k)},\alpha_{j'}^{(k)})\Big]^{\frac{1}{\sigma^{2}_{j'}}}} 
{a(\gamma_{j}^{(k)},\alpha_{j}^{(k)}) b(t_k;\gamma_{j}^{(k)},\alpha_{j}^{(k)}) 
\Big[1 + \tilde{\sigma}^{2}_{j}H^{(k)}_{0j}(t_k,\tilde{\gamma}_{j}^{(k)},\tilde{\alpha}_{j}^{(k)})\Big] \prod\limits_{j'=1}^{L}\Big[1 + \tilde{\sigma}^{2}_{j'}H^{(k)}_{0j'}(t_k,\tilde{\gamma}_{j'}^{(k)},\tilde{\alpha}_{j'}^{(k)})\Big]^ {\frac{1}{\tilde{\sigma}^{2}_{j'}}}},
\end{equation}
for all $t_k>0,\ j=1,\cdots,L$ and $k=1,2$. Then, using similar arguments as those in the proof of \textbf{Theorem} $\boldsymbol{2.1}$, one can show 
\begin{center}
   $\gamma_{j}^{(k)} = \tilde{\gamma}_{j}^{(k)}$ and $\alpha_{j}^{(k)} = \tilde{\alpha}_{j}^{(k)}$ for $j = 1,\cdots,L$ and $k = 1,2$. 
\end{center}

Putting these equalities in \ref{shared_cause_specific_frailty_Eq1}, we get 
\begin{equation}\label{shared_cause_specific_frailty_Eq2}
     \frac{1 + \sigma^{2}_{j}H^{(k)}_{0j}(t_k,\gamma_{j}^{(k)},\alpha_{j}^{(k)})}{1 + \tilde{\sigma}^{2}_{j}H^{(k)}_{0j}(t_k,\gamma_{j}^{(k)},\alpha_{j}^{(k)})} = \frac{\prod\limits_{j'=1}^{L}\Big[1 + \tilde{\sigma}^{2}_{j'}H^{(k)}_{0j'}(t_k,\gamma_{j'}^{(k)},\alpha_{j'}^{(k)})\Big]^{\frac{1}{\tilde{\sigma}^{2}_{j'}}}}{\prod\limits_{j'=1}^{L}\Big[1 + \sigma^{2}_{j'}H^{(k)}_{0j'}(t_k,\gamma_{j'}^{(k)},\alpha_{j'}^{(k)})\Big]^{\frac{1}{\sigma^{2}_{j'}}}} =\frac{S^{(k)}(t_k;\boldsymbol{\xi^{(k)}},\boldsymbol{\tilde{\theta}})}{S^{(k)}(t_k;\boldsymbol{\xi^{(k)}},\boldsymbol{\theta})},
\end{equation}
for all $t_{k} > 0,\ j = 1,\cdots,L$ and $k=1,2$. \\
 
Now, the equality of the joint unconditional sub-distribution functions with $(\boldsymbol{\xi},\boldsymbol{\theta})$ and $(\boldsymbol{\tilde{\xi}},\boldsymbol{\tilde{\theta}})$ also implies equality of the corresponding unconditional survival functions for the $k$th individual; that is, $S^{(k)}(t_k;\boldsymbol{\xi^{(k)}},\boldsymbol{\tilde{\theta}})=S^{(k)}(t_k;\boldsymbol{\xi^{(k)}},\boldsymbol{\theta})$. Using this equality in \ref{shared_cause_specific_frailty_Eq2} yields $\sigma_{j} = \tilde{\sigma}_{j}$, for $j = 1,\cdots,L$. This completes the proof.
\end{proof}

In this case, the cross-ratio function $CR_{j_{1}j_{2}}(t_1,t_2;\boldsymbol{\xi},\sigma_{1},\cdots,\sigma_{L})$ 
between occurrence of failure at time $t_1$ due to cause $j_1$ for the first individual and occurrence of failure at time $t_2$ due to cause $j_2$ for the second individual is given by 
$$\frac{(1 + \sigma^{2}_j) \left(\prod_{k=1}^{2}
h^{(k)}_{0j}(t_k;\boldsymbol{\xi})\right)
S^{2}(t_1,t_2;\boldsymbol{\xi},\boldsymbol{\theta})
    \times\bigg[1 + \sigma^{2}_{j}\sum_{k=1}^2 
    H^{(k)}_{0j}(t_k;\boldsymbol{\xi})\bigg]^{-2}}{\bigg[\int\limits_{t_2}^{\infty}\sum\limits_{j'=1}^{L}f_{jj'}(t_1,u_2;\boldsymbol{\xi},\boldsymbol{\theta})du_2\bigg] \times \bigg[\int\limits_{t_1}^{\infty}\sum\limits_{j'=1}^{L}f_{j'j}(u_1,t_2;\boldsymbol{\xi},\boldsymbol{\theta})du_1\bigg]},$$
for $j_1=j_2=j$, and 
$$\frac{\left(\prod_{k=1}^{2}
h^{(k)}_{0j_k}(t_k;\boldsymbol{\xi})\right)S^{2}(t_1,t_2;\boldsymbol{\xi},\boldsymbol{\theta}) \prod\limits_{k=1}^{2}\bigg[1 + \sigma^{2}_{j_k}\big(H^{(1)}_{0j_k}(t_1;\boldsymbol{\xi}) + H^{(2)}_{0j_k}(t_2;\boldsymbol{\xi})\big)\bigg]^{-1}}{\bigg[\int\limits_{t_2}^{\infty}\sum\limits_{j'=1}^{L}f_{j_1j'}(t_1,u_2;\boldsymbol{\xi},\boldsymbol{\theta})du_2\bigg] \times \bigg[\int\limits_{t_1}^{\infty}\sum\limits_{j'=1}^{L}f_{j'j_2}(u_1,t_2;\boldsymbol{\xi},\boldsymbol{\theta})du_1\bigg]},$$
for $j_1\neq j_2$. \\

\begin{figure}[h]
    \centering
    \subfigure[]{\includegraphics[width=0.45\textwidth]{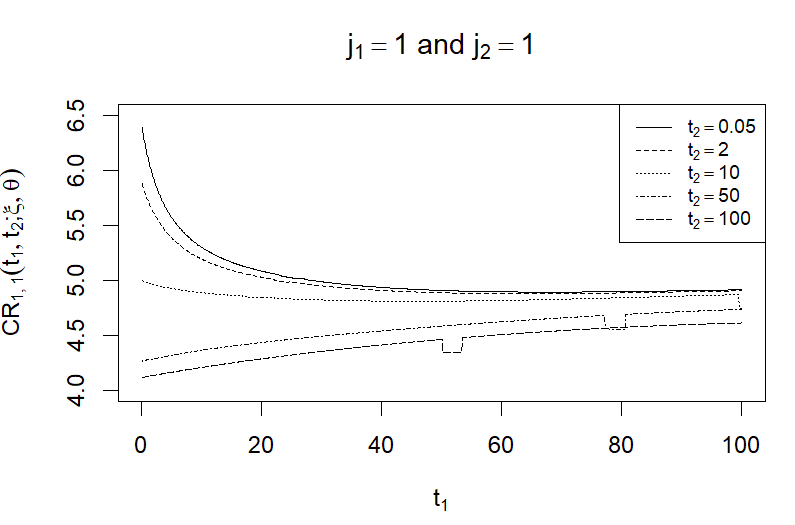}}
    \subfigure[]{\includegraphics[width=0.45\textwidth]{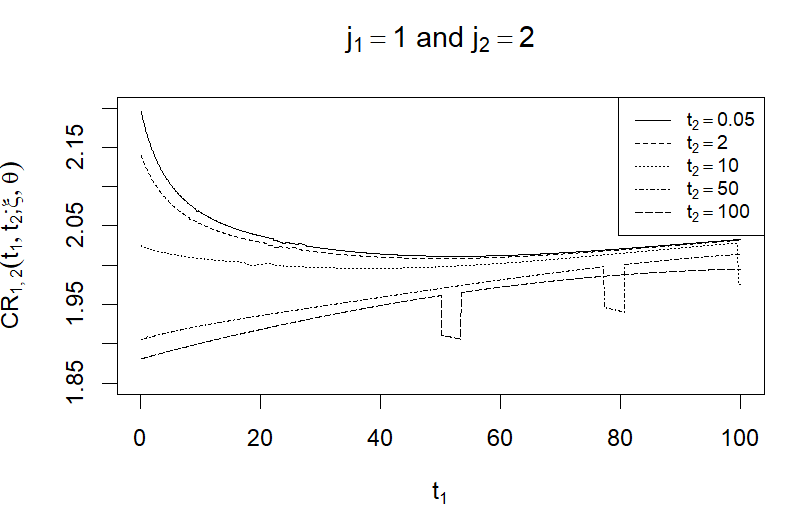}}
    \subfigure[]{\includegraphics[width=0.45\textwidth]{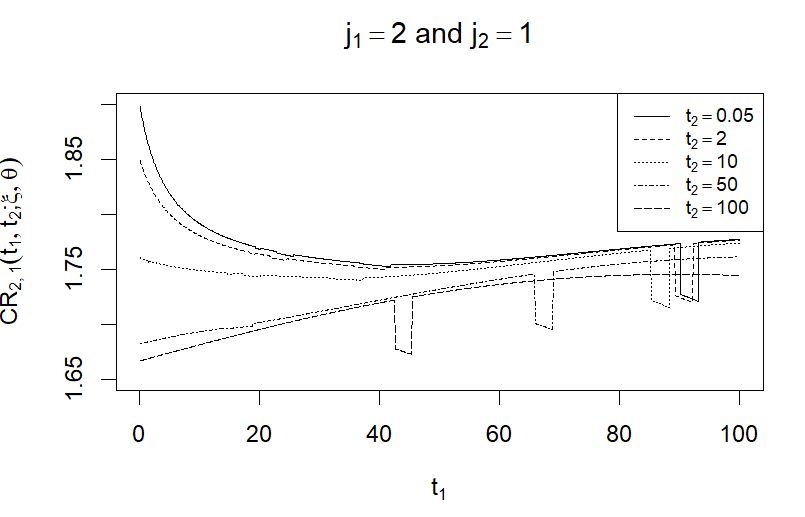}}
    \subfigure[]{\includegraphics[width=0.45\textwidth]{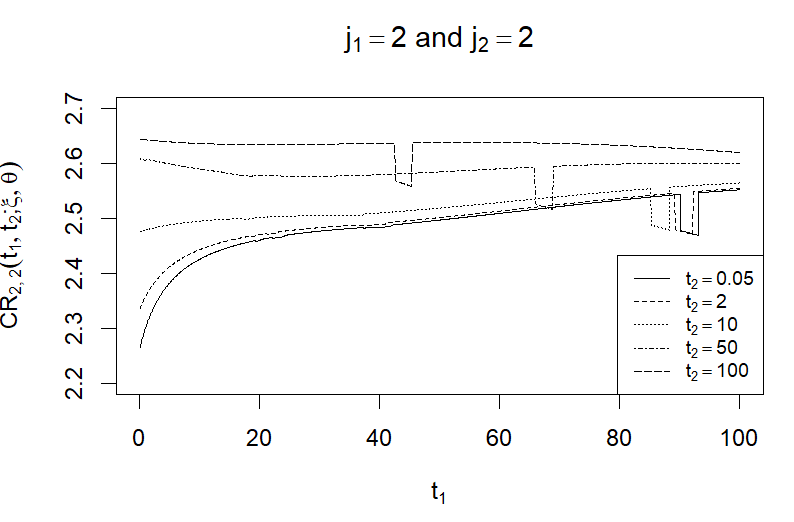}}
    \caption{The cross ratio function corresponding to shared cause specific Gamma frailty}
    \label{fig:Cr-ratio-shared-cause-specific-Gamma}
\end{figure}

It is difficult to find a closed form of this cross-ratio function, even to obtain a suitable lower bound. So, we numerically evaluate the cross-ratio function with two competing risks for different values of $(t_1,t_2)$ and $(j_1,j_2)$, as presented in Figure \ref{fig:Cr-ratio-shared-cause-specific-Gamma}. In this case also, we consider the Exponential type cause-specific hazard rates with the same set of parameters as in Sub-section $2.2$.  
We have taken $(\sigma_1,\sigma_2)$ as $(1.65,0.45)$. We again  plot the cross-ratio function against $t_1$ for the five different values of $t_2$, as in Figure \ref{fig:Cr-ratio-correlated-Gamma}. Note that, in this case, the cross-ratio function depends on $(j_1,j_2)$. The plots for the four different combinations of $(j_1,j_2)$ are presented in the four panels $\ref{fig:Cr-ratio-shared-cause-specific-Gamma}a-\ref{fig:Cr-ratio-shared-cause-specific-Gamma}d$.

We see from Figure \ref{fig:Cr-ratio-shared-cause-specific-Gamma} that  the value of cross-ratio function is greater than unity in each panel. Therefore, failure times of the two individuals due to any particular combination of $(j_1,j_2)$ are positively correlated. 

\subsection{Correlated cause-specific Gamma frailty model}

As in the previous sub-section, the frailty variable depends on the cause of failure but varies with the individual like the correlated frailty. That is, for a particular cause (say, $j$th), there are two different correlated frailty variables $\epsilon^{(1)}_{j}$ and $\epsilon^{(2)}_{j}$ for the first and second individual, respectively, 
depending on that cause $j$, for $j=1,\cdots,L$. Note that, since the frailty variables are cause-specific, the set of competing risks is the same for the two individuals resulting in $L_1=L_2=L$, say. In the example of cancer incidence times of father and son pair, the incidence times may be differently associated for the different cancer sites and 
the possible dependence and unobserved heterogeneity between the father and son is captured through the pair of frailty variables $(\epsilon_j^{(1)},\epsilon_j^{(2)})$ depending on the cause $j$.\\

The correlated cause-specific Gamma frailty model is given by
\begin{equation}\label{correlated cause specific Gamma frailty}
\lambda^{(k)}_{j}\big(t_k;\boldsymbol{\xi}| \epsilon_j^{(k)}\big) = h^{(k)}_{0j}\big(t_k;\boldsymbol{\xi}\big)\epsilon^{(k)}_{j},
\end{equation}
for all $t_{k} > 0$ and $j = 1,\cdots,L,\  k = 1,2$. As in Sub-section 2.2, we assume that the pair $(\epsilon^{(1)}_{j},\epsilon^{(2)}_{j})$ follows a correlated Gamma distribution, as defined therein but with the corresponding set of parameters $(\sigma_{1j},\sigma_{2j},\rho_j)$ 
depending on $j$. We also assume that the $L$ pairs of Gamma frailty variables $(\epsilon_j^{(1)},\epsilon_j^{(2)})$, for $j=1,\cdots,L$, 
are independent. Therefore, the frailty parameter space is  given by
$$\boldsymbol{\Theta} = \{\boldsymbol{\theta}=\left((\sigma_{1j},\sigma_{2j},\rho_j),\ j=1,\cdots,L\right)\},$$
such that $\sigma_{1j} > 0, \sigma_{2j} > 0, 0 < \rho_j < \min\big(\frac{\sigma_{1j}}{\sigma_{2j}},\frac{\sigma_{2j}}{\sigma_{1j}}\big)$, for $j=1,\cdots,L$. As in Sub-section 2.2, the joint density $g(\epsilon_j^{(1)},\epsilon_j^{(2)};\boldsymbol{\theta})$ can be written in terms of three independent Gamma densities of $Y_{0j},\ Y_{1j}$ and $Y_{2j}$ with set of parameters $(\sigma_{1j},\sigma_{2j},\rho_j)$, for $j=1,\cdots,L$. \\ 

We have, conditional on frailty vector $\boldsymbol{\epsilon^{(k)}}=(\epsilon^{(k)}_{1},\cdots,\epsilon^{(k)}_{L})$,  
$$S^{(k)}(t_k;\boldsymbol{\xi}|\boldsymbol{\epsilon^{(k)}}) = \exp{\Big[-\sum\limits_{j=1}^{L}H^{(k)}_{0j}(t_k;\boldsymbol{\xi})\epsilon^{(k)}_{j}\Big]}= 
\exp{\Big[-\sum\limits_{j=1}^{L}H^{(k)}_{0j}(t_k;\boldsymbol{\xi}) \left(\frac{\mu_{0j}}{\mu_{kj}}Y_{0j}+Y_{kj}\right)\Big]}
$$
and 
$$F^{(k)}_{j}(t_k;\boldsymbol{\xi}|\boldsymbol{\epsilon^{(k)}}) 
= \int\limits_{0}^{t_k}h^{(k)}_{0j}\big(u_k;\boldsymbol{\xi}\big)\epsilon^{(k)}_{j}\exp{\Big[-\sum\limits_{j'=1}^{L}H^{(k)}_{0j'}(u_k;\boldsymbol{\xi})\epsilon^{(k)}_{j'}\Big]}du_k,$$
for all $t_{k} > 0,\ j = 1,\cdots,L$ and $k = 1,2$. \\ 

The unconditional joint survival function $S(t_1,t_2;\boldsymbol{\xi},\boldsymbol{\theta})$ is given by
\begin{align*}
    & \prod\limits_{j=1}^{L}\bigg[1 + \sum_{k=1}^2
    \sigma^{2}_{kj}H^{(k)}_{0j}(t_k;\boldsymbol{\xi})\bigg]^{-\frac{\rho_{j}}{\sigma_{1j}\sigma_{2j}}}\times 
    \prod_{k=1}^2
    \prod\limits_{j=1}^{L}\bigg[1 + \sigma^{2}_{kj}H^{(k)}_{0j}(t_k;\boldsymbol{\xi})\bigg]^{\Big(\frac{\rho_{j}}{\sigma_{1j}\sigma_{2j}} - \frac{1}{\sigma^{2}_{kj}}\Big)}\\
    &= \prod\limits_{j=1}^{L}\Big[A_{j}(t_1,t_2;\boldsymbol{\xi},\sigma_{1j},\sigma_{2j})\Big]^{-\frac{\rho_{j}}{\sigma_{1j}\sigma_{2j}}} 
     \prod_{k=1}^{2}\prod\limits_{j=1}^{L}\Big[B_{kj}(t_k;\boldsymbol{\xi},\sigma_{kj})\Big]^{\Big(\frac{\rho_{j}}{\sigma_{1j}\sigma_{2j}} - \frac{1}{\sigma^{2}_{kj}}\Big)}\quad\mbox{ (say),}
\end{align*}
where $A_{j}(t_1,t_2;\boldsymbol{\xi},\sigma_{1j},\sigma_{2j}) = 1 + \sum_{k=1}^2 
\sigma^{2}_{kj}H^{(k)}_{0j}(t_k;\boldsymbol{\xi})$ and 
$B_{kj}(t_k;\boldsymbol{\xi},\sigma_{kj}) = 1 + \sigma^{2}_{kj}H^{(k)}_{0j}(t_k;\boldsymbol{\xi})$, 
for all $t_{k}>0,\ j = 1,\cdots,L$ and $k=1,2$.\\

The unconditional joint sub-distribution function $F_{j_{1}j_{2}}(t_1,t_2;\boldsymbol{\xi},\boldsymbol{\theta})$ is obtained as 
\begin{align*}
&\int\limits_{0}^{t_1}\int\limits_{0}^{t_2} \prod_{k=1}^2\left(\sigma^{2}_{kj}
h^{(k)}_{0j}(u_k;\boldsymbol{\xi})\right) S(u_1,u_2;\boldsymbol{\xi},\boldsymbol{\theta}) \Bigg[\frac{\frac{\rho_{j}}{\sigma_{1j}\sigma_{2j}}\Big(1 + \frac{\rho_{j}}{\sigma_{1j}\sigma_{2j}}\Big)}{\big(A_{j}(u_1,u_2;\boldsymbol{\xi},\sigma_{1j},\sigma_{2j})\big)^2} + 
\prod_{k=1}^2\frac{\frac{1}{\sigma^{2}_{kj}} - \frac{\rho_{j}}{\sigma_{1j}\sigma_{2j}}}
{B_{kj}(u_k;\boldsymbol{\xi},\sigma_{kj})} \\
   &\qquad + \sum_{k=1}^2 
   \frac{\frac{\rho_{j}}{\sigma_{1j}\sigma_{2j}} \left(\frac{1}{\sigma^{2}_{kj}} - \frac{\rho_{j}}{\sigma_{1j}\sigma_{2j}}\right)}
   {A_{j}(u_1,u_2;\boldsymbol{\xi},\sigma_{1j},\sigma_{2j})B_{kj}(u_k;\boldsymbol{\xi},\sigma_{kj})} \Bigg]du_2du_1, 
\end{align*}
for $j_{1} = j_{2} = j$ (say), and 
\begin{align*}
&\int\limits_{0}^{t_1}\int\limits_{0}^{t_2} \prod_{k=1}^2 \left(\sigma^{2}_{kj_k}
h^{(k)}_{0j_k}(u_k;\boldsymbol{\xi})\right) 
S(u_1,u_2;\boldsymbol{\xi},\boldsymbol{\theta})\times 
\Bigg[\prod_{k=1}^2
 \frac{\frac{\rho_{j_k}}{\sigma_{1j_k}\sigma_{2j_k}}}
 {A_{j_k}(u_1,u_2;\boldsymbol{\xi},\sigma_{1j_k},\sigma_{2j_k})} + \prod_{k=1}^2 
   \frac{\frac{1}{\sigma^{2}_{kj_k}} - \frac{\rho_{j_k}}{\sigma_{1j_k}\sigma_{2j_k}}}
   {B_{kj_k}(u_k;\boldsymbol{\xi},\sigma_{kj_k})} + \\ 
 &\qquad \frac{\frac{\rho_{j_2}}{\sigma_{1j_2}\sigma_{2j_2}}\Big(\frac{1}{\sigma^{2}_{1j_1}} - \frac{\rho_{j_1}}{\sigma_{1j_1}\sigma_{2j_1}}\Big)} {\big(A_{j_2}(u_1,u_2;\boldsymbol{\xi},\sigma_{1j_2},\sigma_{2j_2})\big)\big(B_{1j_1}(u_1;\boldsymbol{\xi},\sigma_{1j_1})\big)} 
 + \frac{\frac{\rho_{j_1}}{\sigma_{1j_1}\sigma_{2j_1}}\Big(\frac{1}{\sigma^{2}_{2j_2}} - \frac{\rho_{j_2}}{\sigma_{1j_2}\sigma_{2j_2}}\Big)} {\big(A_{j_1}(u_1,u_2;\boldsymbol{\xi},\sigma_{1j_1},\sigma_{2j_1})\big)\big(B_{2j_2}(u_2;\boldsymbol{\xi},\sigma_{2j_2})\big)} \Bigg]du_2du_1,
\end{align*}
for $j_{1} \neq j_{2}$. 
Accordingly, the unconditional joint sub-density function $f_{j_{1}j_{2}}(t_1,t_2;\boldsymbol{\xi},\boldsymbol{\theta})$ is given by 
\begin{align*}
&\prod_{k=1}^2\left(\sigma^{2}_{kj}
h^{(k)}_{0j}(t_k;\boldsymbol{\xi})\right) S(t_1,t_2;\boldsymbol{\xi},\boldsymbol{\theta}) \Bigg[\frac{\frac{\rho_{j}}{\sigma_{1j}\sigma_{2j}}\Big(1 + \frac{\rho_{j}}{\sigma_{1j}\sigma_{2j}}\Big)}{\big(A_{j}(t_1,t_2;\boldsymbol{\xi},\sigma_{1j},\sigma_{2j})\big)^2} + 
\prod_{k=1}^2\frac{\frac{1}{\sigma^{2}_{kj}} - \frac{\rho_{j}}{\sigma_{1j}\sigma_{2j}}}
{B_{kj}(t_k;\boldsymbol{\xi},\sigma_{kj})} \\
   &\qquad + \sum_{k=1}^2 
   \frac{\frac{\rho_{j}}{\sigma_{1j}\sigma_{2j}} \left(\frac{1}{\sigma^{2}_{kj}} - \frac{\rho_{j}}{\sigma_{1j}\sigma_{2j}}\right)}
   {A_{j}(t_1,t_2;\boldsymbol{\xi},\sigma_{1j},\sigma_{2j})B_{kj}(t_k;\boldsymbol{\xi},\sigma_{kj})} \Bigg],  
\end{align*}
for $j_{1} = j_{2} = j$ (say), and 
\begin{align*}
 & \prod_{k=1}^2 \left(\sigma^{2}_{kj_k}h^{(k)}_{0j_k}(t_k;\boldsymbol{\xi})\right) S(t_1,t_2;\boldsymbol{\xi},\boldsymbol{\theta})\times 
 \Bigg[\prod_{k=1}^2 \frac{\frac{\rho_{j_k}}{\sigma_{1j_k}\sigma_{2j_k}}}
 {A_{j_k}(t_1,t_2;\boldsymbol{\xi},\sigma_{1j_k},\sigma_{2j_k})} 
 + \prod_{k=1}^2
   \frac{\frac{1}{\sigma^{2}_{kj_k}} - \frac{\rho_{j_k}}{\sigma_{1j_k}\sigma_{2j_k}}}
   {B_{kj_k}(t_k;\boldsymbol{\xi},\sigma_{kj_k})} \\
 &\qquad 
 + \frac{\frac{\rho_{j_2}}{\sigma_{1j_2}\sigma_{2j_2}}\Big(\frac{1}{\sigma^{2}_{1j_1}} - \frac{\rho_{j_1}}{\sigma_{1j_1}\sigma_{2j_1}}\Big)}{A_{j_2}(t_1,t_2;\boldsymbol{\xi},\sigma_{1j_2},\sigma_{2j_2})B_{1j_1}(t_1;\boldsymbol{\xi},\sigma_{1j_1})}
    + \frac{\frac{\rho_{j_1}}{\sigma_{1j_1}\sigma_{2j_1}}\Big(\frac{1}{\sigma^{2}_{2j_2}} - \frac{\rho_{j_2}}{\sigma_{1j_2}\sigma_{2j_2}}\Big)}
   {A_{j_1}(t_1,t_2;\boldsymbol{\xi},\sigma_{1j_1},\sigma_{2j_1})B_{2j_2}(t_2;\boldsymbol{\xi},\sigma_{2j_2})} \Bigg],
    \end{align*}
for $j_{1} \neq j_{2}$. \\

\begin{definition}
    The correlated cause-specific Gamma frailty model $(\ref{correlated cause specific Gamma frailty})$ is identifiable within the family  $\boldsymbol{\Xi}\times\boldsymbol{\Theta}$ with $\boldsymbol{\Theta}$ as described above in this sub-section 
    if, for some $\boldsymbol{\xi},\boldsymbol{\tilde{\xi}} \in \boldsymbol{\Xi}$ and $\boldsymbol{\theta}=((\sigma_{1j},\sigma_{2j},\rho_j),\ j=1,\cdots,L),\ \boldsymbol{\tilde{\theta}}=((\tilde{\sigma}_{1j},\tilde{\sigma}_{2j},\tilde{\rho}_j),\ j=1,\cdots,L) \in \boldsymbol{\Theta}$,  the equality 
 $F_{j_{1}j_{2}}(t_1,t_2;\boldsymbol{\xi},\boldsymbol{\theta}) = F_{j_{1}j_{2}}(t_1,t_2;\boldsymbol{\tilde{\xi}},\boldsymbol{\tilde{\theta}})$ with the above expressions, 
for all $t_{k} > 0, j_{k} = 1,\cdots,L$ and $k = 1,2$, implies 
$\boldsymbol{\xi}=\boldsymbol{\tilde{\xi}}$ and $\sigma_{kj} = \tilde{\sigma}_{kj},\ \rho_j=\tilde{\rho}_j$ for $j = 1,\cdots,L$ and $k = 1,2$. 
\end{definition}

As in the previous sub-setions, the unconditional $j$th sub-distribution function $F^{(k)}_{j}(t_k;\boldsymbol{\xi^{(k)}},\sigma_{kj})$ 
for the $k$th individual depends only on $\boldsymbol{\xi^{(k)}}$ and $\sigma_{kj}$ and is given by the same expression as that in $(\ref{marg_subdis_shared_Gamma_frailty})$, with $\sigma$ replaced by $\sigma_{kj}$, for $j = 1,\cdots,L$ and $k = 1,2$. \\

\begin{theorem}
   The correlated cause-specific Gamma frailty model (\ref{correlated cause specific Gamma frailty}), with the cause-specific baseline hazard functions belonging to the family defined in (\ref{parametric_family_haz}), is identifiable within $\boldsymbol{\Xi}\times\boldsymbol{\Theta}$. 
\end{theorem}

\begin{proof}
Following the same steps as in the proof of \textbf{Theorem} $\mathbf{2.3}$, using the equality of the unconditional marginal sub-density functions $f^{(k)}_{j}(t_k;\boldsymbol{\xi^{(k)}},\sigma_{kj})$ and $f^{(k)}_{j}(t_k;\boldsymbol{\tilde{\xi}^{(k)}},\tilde{\sigma}_{kj})$, one can prove that 
$$\xi^{(k)}_{j} = \tilde{\xi}^{(k)}_{j} \, \, \text{and} \, \, \sigma_{kj} = \tilde{\sigma}_{kj},$$
for all $j = 1,\cdots,L$ and $k = 1,2$. With these equalities and the equality of unconditional joint sub-density functions $f_{jj}(t_1,t_2;\boldsymbol{\xi},\boldsymbol{\theta})$ and $f_{jj}(t_1,t_2;\boldsymbol{\tilde{\xi}},\boldsymbol{\tilde{\theta}})$, for $t_{k} > 0,\  j = 1,\cdots,L$ and  $k = 1,2$, as implied by \textbf{Definition} $\mathbf{5}$, we get  
\begin{align*} 
&\Bigg[\frac{\frac{\rho_{j}}{\sigma_{1j}\sigma_{2j}}\Big(1 + \frac{\rho_{j}}{\sigma_{1j}\sigma_{2j}}\Big)}{\big(A_{j}(t_1,t_2;\boldsymbol{\xi},\sigma_{1j},\sigma_{2j})\big)^2} + 
\sum_{k=1}^2 
\frac{\frac{\rho_{j}}{\sigma_{1j}\sigma_{2j}}(\frac{\rho_{j}}{\sigma_{1j}\sigma_{2j}} - \frac{1}{\sigma^{2}_{kj}})}{A_{j}(t_1,t_2;\boldsymbol{\xi},\sigma_{1j},\sigma_{2j})B_{kj}(t_k;\boldsymbol{\xi},\sigma_{kj})} 
+ \prod_{k=1}^2 \frac{(\frac{\rho_{j}}{\sigma_{1j}\sigma_{2j}} - \frac{1}{\sigma^{2}_{kj}})}
{B_{kj}(t_k;\boldsymbol{\xi},\sigma_{kj})}\Bigg]\times  \\
& \prod\limits_{j'=1}^{L}\Big[A_{j'}(t_1,t_2;\boldsymbol{\xi},\sigma_{1j'},\sigma_{2j'})\Big]^{-\Big(\frac{\rho_{j'}}{\sigma_{1j'}\sigma_{2j'}}\Big)}
\prod_{k=1}^2\prod\limits_{j'=1}^{L}\Big[B_{kj'}(t_k;\boldsymbol{\xi},\sigma_{1j'})\Big]^{\Big(\frac{\rho_{j'}}{\sigma_{1j'}\sigma_{2j'}} - \frac{1}{\sigma^{2}_{kj'}}\Big)} = \\
&\Bigg[\frac{\frac{\tilde{\rho}_{j}}{\sigma_{1j}\sigma_{2j}}\Big(1 + \frac{\tilde{\rho}_{j}}{\sigma_{1j}\sigma_{2j}}\Big)}{\big(A_{j}(t_1,t_2;\boldsymbol{\xi},\sigma_{1j},\sigma_{2j})\big)^2} + 
\sum_{k=1}^2 
\frac{\frac{\tilde{\rho}_{j}}{\sigma_{1j}\sigma_{2j}}(\frac{\tilde{\rho}_{j}}{\sigma_{1j}\sigma_{2j}} - \frac{1}{\sigma^{2}_{kj}})}{A_{j}(t_1,t_2;\boldsymbol{\xi},\sigma_{1j},\sigma_{2j})B_{kj}(t_k;\boldsymbol{\xi},\sigma_{kj})} 
+ \prod_{k=1}^2 \frac{(\frac{\tilde{\rho}_{j}}{\sigma_{1j}\sigma_{2j}} - \frac{1}{\sigma^{2}_{kj}})}
{B_{kj}(t_k;\boldsymbol{\xi},\sigma_{kj})}\Bigg]\times  \\
& \prod\limits_{j'=1}^{L}\Big[A_{j'}(t_1,t_2;\boldsymbol{\xi},\sigma_{1j'},\sigma_{2j'})\Big]^{-\Big(\frac{\tilde{\rho}_{j'}}{\sigma_{1j'}\sigma_{2j'}}\Big)}
\prod_{k=1}^2\prod\limits_{j'=1}^{L}\Big[B_{kj'}(t_k;\boldsymbol{\xi},\sigma_{1j'})\Big]^{\Big(\frac{\tilde{\rho}_{j'}}{\sigma_{1j'}\sigma_{2j'}} - \frac{1}{\sigma^{2}_{kj'}}\Big)}
\end{align*}
for $t_{k} > 0, j = 1,\cdots,L$ and $k = 1,2$. \\

Note that, as $t_{k} \to 0+$, for $k=1,2$, we have 
$A_{j}(t_1,t_2;\boldsymbol{\xi},\sigma_{1j},\sigma_{2j}) \to 1$, $B_{kj}(t_k;\boldsymbol{\xi}, \sigma_{1j}) \to 1$, for $k=1,2$. Hence, letting $t_{k} \to 0+$, for $k=1,2$, in the above equality, we get 
\begin{multline*}
    \frac{\rho_{j}}{\sigma_{1j}\sigma_{2j}}\Big(1 + \frac{\rho_{j}}{\sigma_{1j}\sigma_{2j}}\Big) + \frac{\rho_{j}}{\sigma_{1j}\sigma_{2j}}\Big(\frac{1}{\sigma^{2}_{1j}} - \frac{\rho_{j}}{\sigma_{1j}\sigma_{2j}}\Big) + \frac{\rho_{j}}{\sigma_{1j}\sigma_{2j}}\Big(\frac{1}{\sigma^{2}_{2j}} - \frac{\rho_{j}}{\sigma_{1j}\sigma_{2j}}\Big) + \Big(\frac{1}{\sigma^{2}_{1j}} - \frac{\rho_{j}}{\sigma_{1j}\sigma_{2j}}  \Big)\Big(\frac{1}{\sigma^{2}_{2j}} - \frac{\rho_{j}}{\sigma_{1j}\sigma_{2j}}\Big)\\
    = 
\frac{\tilde{\rho}_{j}}{\sigma_{1j}\sigma_{2j}}\Big(1 + \frac{\tilde{\rho}_{j}}{\sigma_{1j}\sigma_{2j}}\Big) + \frac{\tilde{\rho}_{j}}{\sigma_{1j}\sigma_{2j}}\Big(\frac{1}{\sigma^{2}_{1j}} - \frac{\tilde{\rho}_{j}}{\sigma_{1j}\sigma_{2j}}\Big) + \frac{\tilde{\rho}_{j}}{\sigma_{1j}\sigma_{2j}}\Big(\frac{1}{\sigma^{2}_{2j}} - \frac{\tilde{\rho}_{j}}{\sigma_{1j}\sigma_{2j}}\Big) + \Big(\frac{1}{\sigma^{2}_{1j}} - \frac{\tilde{\rho}_{j}}{\sigma_{1j}\sigma_{2j}}\Big)\Big(\frac{1}{\sigma^{2}_{2j}} - \frac{\tilde{\rho}_{j}}{\sigma_{1j}\sigma_{2j}}\Big)
\end{multline*}
for $j = 1,\cdots,L$. 
Therefore, both sides of the above equality being the same quadratic functions of in $\rho_{j}$ and $\tilde{\rho}_{j}$, respectively, 
we get $\rho_{j} = \tilde{\rho}_{j}$ for $j = 1,\cdots,L$. This completes the proof.
\end{proof}

The cross ratio function $CR_{j_{1}j_{2}}(t_1,t_2;\boldsymbol{\xi},\boldsymbol{\theta})$ between the occurrence of failure at time $t_1$ due to cause $j_1$ for the first individual and occurrence of failure at time $t_2$ due to cause $j_2$ for the second individual is 
\begin{align*}
& \Bigg[\frac{\prod_{k=1}^2 \left(
\sigma^{2}_{kj}h^{(k)}_{0j}(t_k;\boldsymbol{\xi})\right)
\left(S(t_1,t_2;\boldsymbol{\xi},\boldsymbol{\theta})\right)^{2}}{\left(\int\limits_{t_2}^{\infty} \sum\limits_{j'=1}^{L}f_{jj'}(t_1,u_2;\boldsymbol{\xi},\boldsymbol{\theta})du_2\right) \times\left( \int\limits_{t_1}^{\infty} \sum\limits_{j'=1}^{L}f_{j'j}(u_1,t_2;\boldsymbol{\xi},\boldsymbol{\theta})du_1\right)} \Bigg]\times \\
    &\qquad \Bigg[\frac{\frac{\rho_{j}}{\sigma_{1j}\sigma_{2j}}\Big(1 + \frac{\rho_{j}}{\sigma_{1j}\sigma_{2j}}\Big)}{\big(A_{j}(t_1,t_2;\boldsymbol{\xi},\sigma_{1j},\sigma_{2j})\big)^2} + \sum_{k=1}^2 \frac{\frac{\rho_{j}}{\sigma_{1j}\sigma_{2j}}\Big(\frac{\rho_{j}}{\sigma_{1j}\sigma_{2j}} - \frac{1}{\sigma^{2}_{kj}}\Big)}{A_{j}(t_1,t_2;\boldsymbol{\xi},\sigma_{1j},\sigma_{2j})B_{kj}(t_k;\boldsymbol{\xi},\sigma_{kj})}+ \prod_{k=1}^2 
     \frac{(\frac{\rho_{j}}{\sigma_{1j}\sigma_{2j}} - \frac{1}{\sigma^{2}_{kj}})}
     {B_{kj}(t_k;\boldsymbol{\xi},\sigma_{kj})}\Bigg],
\end{align*}
for $j_1=j_2=j$, say, and 
\begin{align*}
&\Bigg[\frac{\prod_{k=1}^2\left(\sigma^{2}_{kj_k} h^{(k)}_{0j_k}(t_k;\boldsymbol{\xi})\right)\left( S(t_1,t_2;\boldsymbol{\xi},\boldsymbol{\theta})\right)^{2}}{\left(\int\limits_{t_2}^{\infty} \sum\limits_{j=1}^{L} f_{j_1j}(t_1,u_2;\boldsymbol{\xi},\boldsymbol{\theta}) du_2\right) \times\left(\int\limits_{t_1}^{\infty} \sum\limits_{j=1}^{L}f_{jj_2}(u_1,t_2;\boldsymbol{\xi},\boldsymbol{\theta})du_1\right)}\Bigg]\times \Bigg[\prod_{k=1}^2 
 \frac{\frac{\rho_{j_k}}{\sigma_{1,j_k}\sigma_{2,j_k}}}{A_{j_k}(t_1,t_2;\boldsymbol{\xi},\sigma_{1j_k},\sigma_{2j_k})} + \\ 
    & \frac{\frac{\rho_{j_2}}{\sigma_{1j_2}\sigma_{2j_2}}\Big(\frac{\rho_{j_1}}{\sigma_{1j_1}\sigma_{2j_1}} - \frac{1}{\sigma^{2}_{1j_1}}\Big)}{A_{j_2}(t_1,t_2;\boldsymbol{\xi},\sigma_{1j_2},\sigma_{2j_2})B_{1j_1}(t_1;\boldsymbol{\xi},\sigma_{1j_1})} 
   + \frac{\frac{\rho_{j_1}}{\sigma_{1j_1}\sigma_{2j_1}}\Big(\frac{\rho_{j_2}}{\sigma_{1j_2}\sigma_{2j_2}} - \frac{1}{\sigma^{2}_{2j_2}}\Big)}{A_{j_1}(t_1,t_2;\boldsymbol{\xi},\sigma_{1j_1},\sigma_{2j_1})B_{2j_2}(t_2;\boldsymbol{\xi},\sigma_{2j_2})} +\prod_{k=1}^2 
   \frac{\Big(\frac{\rho_{j_k}}{\sigma_{1j_k}\sigma_{2j_k}} - \frac{1}{\sigma^{2}_{kj_k}}\Big)}
   {B_{kj_k}(t_k;\boldsymbol{\xi},\sigma_{kj_k})}\Bigg], 
\end{align*}
for $j_1\ne j_2$. \\

\begin{figure}[h]
    \centering
    \subfigure[]{\includegraphics[width=0.45\textwidth]{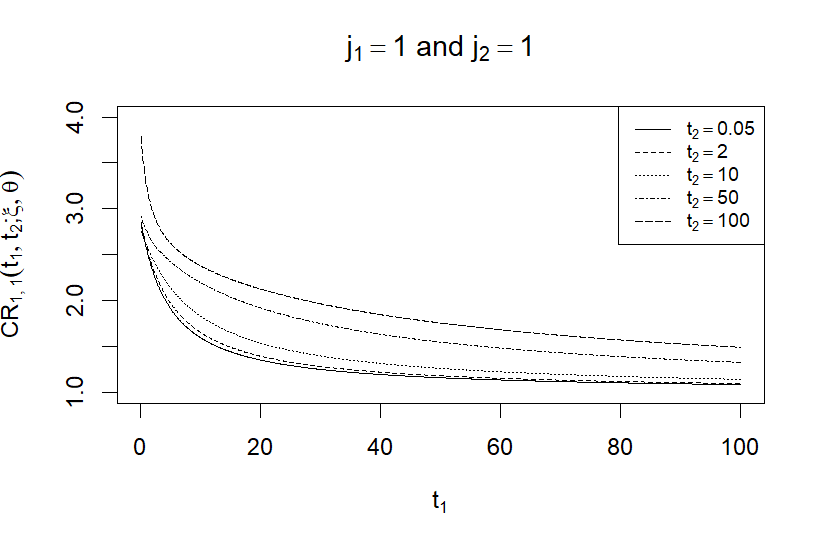}}
    \subfigure[]{\includegraphics[width=0.45\textwidth]{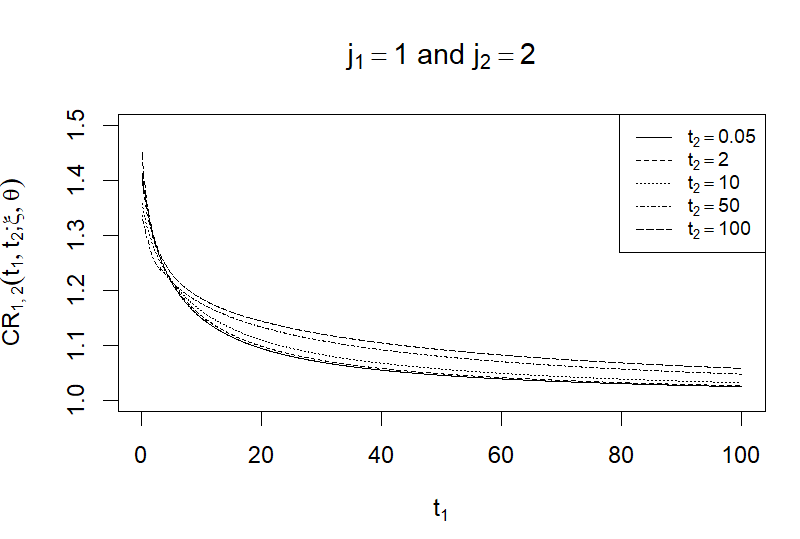}}
    \subfigure[]{\includegraphics[width=0.45\textwidth]{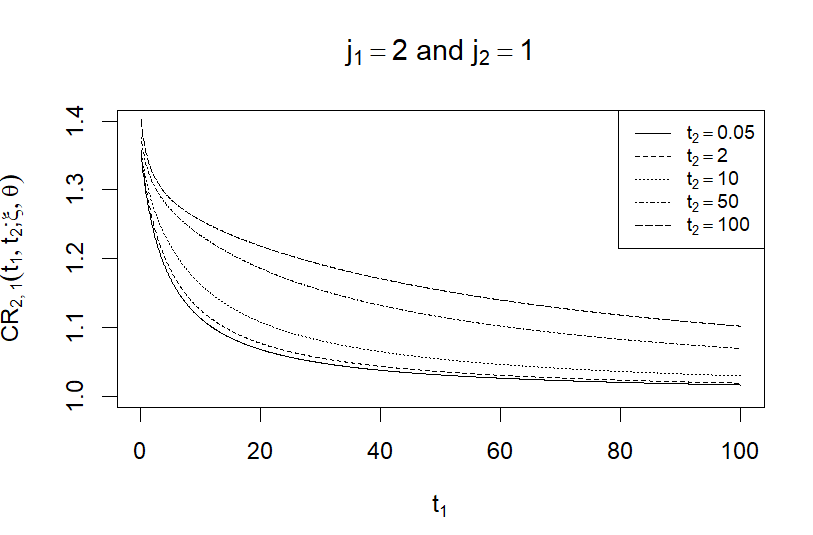}}
    \subfigure[]{\includegraphics[width=0.45\textwidth]{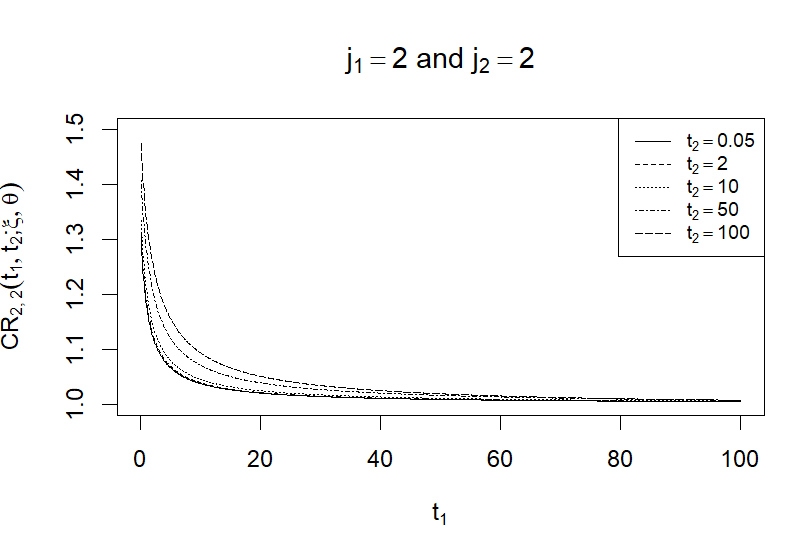}}
    \caption{The cross ratio function corresponding to correlated cause specific Gamma frailty}
    \label{fig:Cr-ratio-correlated-cause-specific-Gamma}
\end{figure}

As in the previous sub-section, it is also difficult to find a closed form of this cross-ratio, even to obtain a suitable lower bound. So, we numerically evaluate the cross-ratio function for different values of $(t_1,t_2)$ and $(j_1,j_2)$, as presented in Figure \ref{fig:Cr-ratio-correlated-cause-specific-Gamma}. In this case also, we consider the Exponential type cause-specific hazard rates with the same set of parameters as before. Also, we take $(\sigma_{11},\sigma_{12})=(1.2,1.8)$ for the first individual and $(\sigma_{21},\sigma_{22})=(0.8,0.4)$ for  the second individual. The cause-specific correlation parameters $\rho_1$ and $\rho_2$ are taken as 0.7 and 0.25, respectively. As before, we plot the cross-ratio function against $t_1$ for the five different values of $t_2$ in the four panels $\ref{fig:Cr-ratio-correlated-cause-specific-Gamma}a-\ref{fig:Cr-ratio-correlated-cause-specific-Gamma}d$ for the four different combinations of $(j_1,j_2)$. \\

Again, we see from Figure \ref{fig:Cr-ratio-correlated-cause-specific-Gamma} that  the value of cross-ratio function is greater than unity in each panel. Therefore, failure times of the two individuals due to any particular combination of $(j_1,j_2)$ are positively correlated. 

\vskip10pt

\section{Maximum Likelihood Estimation}

\vskip10pt

Recall, from the beginning of Section 2, that the observed data is of the form $(X_1,X_2,J_1,J_2)$. Let us write $J_k=0$ when the failure time corresponding to the $k$th individual is right censored at the monitoring time $X_k$, for $k=1,2$. A typical observation from a pair of individuals is, therefore, $(x_1,x_2,j_1,j_2)$, where $x_1>0$ and $x_2>0$ denote the monitoring times of the two individuals with $j_1$ and $j_2$ being the corresponding causes of failure, for $j_k=0,1,\cdots,L_k,\ k=1,2$. 
In order to construct the likelihood function based on observations $(x_1,x_2,j_1,j_2)$ from $n$ independent pairs, let us consider likelihood contributions for different combinations of $(j_1,j_2)$, as given below. \\

Let $h(x_1,x_2)$ denote the joint density of the two monitoring times $(X_1,X_2)$, which is not of interest and considered as nuisance. Then, for $j_k=1,\cdots,L_k,\ k=1,2$, the likelihood contribution is proportional to $P[T_{1} \leq X_1=x_1,T_{2} \leq X_2=x_2,J_1 = j_1,J_2 = j_2;\boldsymbol{\xi},\boldsymbol{\theta}]h(x_1,x_2)$, which is given by 
$$f^{\ast}(x_1,x_2,j_1,j_2;\boldsymbol{\xi},\boldsymbol{\theta})= \Bigg[ \int\limits_{0}^{x_1} \int\limits_{0}^{x_2}f_{j_{1}j_{2}}(u_1,u_2;\boldsymbol{\xi},\boldsymbol{\theta}) du_{1}du_{2}\Bigg]h(x_1,x_2)\propto F_{j_{1}j_{2}}(x_{1},x_{2};\boldsymbol{\xi},\boldsymbol{\theta}). $$

Also, for $j_1 = 0$ and $j_2 \in \{1,\cdots,L_2\}$, we have $\{T_{1} > X_1=x_1,\ T_{2} \leq X_2=x_2, J_2=j_2\}$. Then, 
the likelihood contribution is proportional to  

\begin{eqnarray*} 
f^{\ast}(x_1,x_2,j_1,j_2;\boldsymbol{\xi},\boldsymbol{\theta}) &=&  \Bigg[\int\limits_{0}^{x_2} f^{(2)}_{j_2}(u_2;\boldsymbol{\xi},\boldsymbol{\theta})du_{2} - \int\limits_{0}^{x_1}\int\limits_{0}^{x_2}\sum_{j=1}^{L_1}f_{jj_2}(u_1,u_2;\boldsymbol{\xi},\boldsymbol{\theta}) du_{1}du_{2}\Bigg]h(x_1,x_2) \\
&\propto & F_{j_2}^{(2)}(x_{2};\boldsymbol{\xi},\boldsymbol{\theta}) - \sum\limits_{j=1}^{L_1}F_{jj_2}(x_{1},x_{2};\boldsymbol{\xi},\boldsymbol{\theta}). 
\end{eqnarray*}

Similarly, for $j_1 \in \{1,\cdots,L_1\}$ and $j_2 = 0$, we have $\{T_{1} \leq X_1=x_1, J_1=j_1,\ T_{2} > X_2=x_2\}$ and 
the likelihood contribution is proportional to 
\begin{eqnarray*} 
f^{\ast}(x_1,x_2,j_1,j_2;\boldsymbol{\xi},\boldsymbol{\theta}) &=& \Bigg[\int\limits_{0}^{x_1} f^{(1)}_{j_1}(u_1;\boldsymbol{\xi},\boldsymbol{\theta})du_{1} - \int\limits_{0}^{x_1}\int\limits_{0}^{x_2}\sum_{j=1}^{L_2}f_{j_1j}(u_1,u_2;\boldsymbol{\xi},\boldsymbol{\theta}) du_{1}du_{2}\Bigg]h(x_1,x_2)\\ 
&\propto & F_{j_1}^{(1)}(x_{1};\boldsymbol{\xi},\boldsymbol{\theta}) - \sum\limits_{j=1}^{L_2}F_{j_1j}(x_{1},x_{2};\boldsymbol{\xi},\boldsymbol{\theta}). 
\end{eqnarray*}

Finally, when failure times of both the individuals are censored (that is, $j_1=j_2=0$), we have $\{T_{1}> X_1=x_1,\ T_{2} > X_2=x_2\}$ and 
the likelihood contribution is proportional to 
$$f^{\ast}(x_1,x_2,j_1,j_2;\boldsymbol{\xi},\boldsymbol{\theta})= \Bigg[\int\limits_{x_1}^{\infty}\int\limits_{x_2}^{\infty}f(u_1,u_2;\boldsymbol{\xi},\boldsymbol{\theta}) du_{1}du_{2}\Bigg]h(x_1,x_2) \propto S(x_{1},x_{2};\boldsymbol{\xi},\boldsymbol{\theta}).$$

Suppose we have data from $n$ independent pairs of individuals giving observations $(x_{i1},x_{i2},j_{i1},j_{i2})$, for $i=1,2,\cdots,n$. Then, the likelihood function is given by 
\begin{align*}
    L(\boldsymbol{\xi,\boldsymbol{\theta}}) &\propto \prod\limits_{i=1}^{n}f^{\ast}(x_{i1},x_{i2},j_{i1},j_{i2};\boldsymbol{\xi},\boldsymbol{\theta}).
\end{align*}

It is important to note that we have a fairly complex likelihood function so that it is not possible to obtain closed form solution of the maximum likelihood estimators (MLEs) of the model parameters. Therefore, we have to make use of some efficient numerical maximization procedure to compute the MLEs. We have performed the task of numerical maximization of the likelihood function with the help of parallel computing. We have used \textbf{optimParallel} function from the \textbf{optimParallel} package of \textbf{R} statistical software for this purpose. The \textbf{optimParallel} function generally requires a cluster of $1+2p$ processor cores to solve a minimization problem consisting of $p$ parameters. The computation burden is lesser for the shared and correlated Gamma frailty models, because we have closed form expressions  of the joint sub-distribution functions. 
To speed up the parallel computation in case of shared and correlated cause-specific Gamma frailty models, we have also used \textbf{RcppNumerical} package of \textbf{R} software to numerically evaluate the integrals present in the likelihood functions. The asymptotic variance-covariance matrix of the MLEs is estimated by inverting the corresponding hessian matrix computed at the MLEs. \\

let $\boldsymbol{\eta_{0}}=(\boldsymbol{\xi_0},\boldsymbol{\theta_0})$ be the true parameter vector. 
Since the likelihood function can be expressed as a product of $n$ independent and identically distributed (iid) densities $f^{\ast}(x_{i1},x_{i2},j_{i1},j_{i2};\boldsymbol{\xi},\boldsymbol{\theta})$'s, derivation of the asymptotic results, including consistency and asymptotic normality of maximum likelihood estimator $\boldsymbol{\hat{\eta}} = (\boldsymbol{\hat{\xi}},\boldsymbol{\hat{\theta}})$ of, follow routinely from the standard likelihood theory under some regularity conditions. The regularity conditions required for the asymptotic results are: 
\begin{enumerate}
    \item The parameter vector $\boldsymbol{\eta}$ is identifiable with respect to the density function $f^{\ast}(.;\boldsymbol{\eta})$.
    \item The parameter space $\boldsymbol{\Xi}\times\boldsymbol{\Theta}$ is open. 
    \item The observations $(x_{i1},x_{i2},j_{i1},j_{i2})$, for $i = 1,2,\cdots,n$, are iid random vectors with common density $f^{\ast}(.;\boldsymbol{\eta})$. 
    \item The sample space of the random vector $(X_{1},X_{2},J_{1},J_{2})$, that is the support of the density $f^{\ast}(.;\boldsymbol{\eta})$, is independent of the parameter vector $\boldsymbol{\eta}$. 
    \item We assume that all possible third-order partial derivatives of $f^{\ast}(.;\boldsymbol{\eta})$ exist and are continuous functions of $\boldsymbol{\eta}$.
    \item The third-order partial derivatives of $f^{\ast}(.;\boldsymbol{\eta})$ are bounded above by functions of $(x_{1},x_2,j_1,j_2)$ in the neighbourhood of the true parameter vector $\boldsymbol{\eta_{0}}$.
    \item The Fisher information matrix corresponding to the joint density $f^{\ast}(.;\boldsymbol{\eta})$ is assumed to be positive definite.
\end{enumerate}

Then, we have the following theorem, the proof of which follows from Lehmann and Casella (1998, p449) and hence skipped. 

\begin{theorem}
    The maximum likelihood estimator of the parameter vector $\boldsymbol{\eta}$, denoted by $\boldsymbol{\hat{\eta}}$, satisfies
    \begin{enumerate}
        \item $\boldsymbol{\hat{\eta}} \overset{\text{P}}{\to} \boldsymbol{\eta_0}$,
        \item $\sqrt{n}(\boldsymbol{\hat{\eta}} - \boldsymbol{\eta_0})$ asymptotically follows a mean zero multivariate Normal distribution with variance-covariance matrix estimated by the inverse of hessian matrix computed at $\boldsymbol{\hat{\eta}}$.
    \end{enumerate}
\end{theorem}

\vskip10pt

\section{Simulation Studies}

\vskip10pt

In order to carry out a simulation study to investigate the finite sample behaviour of $\boldsymbol{\hat{\eta}}$, we consider the same monitoring time $X=x$ for both the individuals in a pair, which vary over the different pairs according to a distribution given by the density function $h(x;\mu)$, say, with an associated  parameter $\mu$. For our study, we consider Exponential  distribution with mean $1/\mu$ for the common monitoring time $X$. 
In order to study the effect of censoring on the estimate $\boldsymbol{\hat{\eta}}$, we choose the  value of $\mu$ so that the probability of both the failure times $T_1$ and $T_2$ being censored equals a fixed value $p_{cen}$, say. That is, 
$$\int\limits_{0}^{\infty}P[T_1 > x,T_2 > x;\boldsymbol{\eta_0}]h(x;\mu)dx = p_{cen}.$$ 
We consider different values of $p_{cen}$ resulting in different values of $\mu$ by numerically solving the above equation. \\

For a sample size of $n$ pairs, we simulate $n$ iid monitoring times $x_1,\cdots,x_n$ from the Exponential  distribution with mean $1/\mu$. Note that, for each $i$, given $X_1=X_2=x_i$, the pair $(J_1,J_2)$ follows a multinomial distribution with $(L_1 + 1) \times (L_2 + 1)$ number of cells and the cell probabilities as worked out in Section 3. Therefore, given the monitoring time $x_i$, we simulate the cause of failure $J_k=j_{ik}$, taking values in $\{0,1,\cdots,L_k\}$, for $k=1,2$, using the above multinomial distribution. Repeating this for  $i=1,\cdots,n$, we have one simulated sample of bivariate current status data with competing risks, $\{(x_i,x_i,j_{i1},j_{i2});i = 1,\cdots,n\}$. The above procedure is repeated $10000$ times to get $10000$ such simulated samples each of size $n$.\\ 

For our simulation, we consider two competing risks for each individual ($L_1=L_2=2$) with constant or Exponential type  baseline cause-specific hazard function for each cause; that is, $h_{0j}(t_k;\boldsymbol{\xi})=\alpha_j^{(k)}$, for $j=1,2$ and $k=1,2$, with $\boldsymbol{\xi}= 
(\alpha^{(1)}_{1},\alpha^{(1)}_{2},\alpha^{(2)}_{1},\alpha^{(2)}_{2})$. We consider the four different frailty models discussed in Section 2. For the shared frailty model, we take $(\alpha^{(1)}_{1},\alpha^{(1)}_{2},\alpha^{(2)}_{1},\alpha^{(2)}_{2}) = (2.4,5.8,3.5,4.5)$ and $\boldsymbol{\theta}=\sigma=0.95$. Similarly, for the correlated Gamma frailty model, we take 
$(\alpha^{(1)}_{1},\alpha^{(1)}_{2},\alpha^{(2)}_{1},\alpha^{(2)}_{2}) = (3,2,2.5,0.8)$ and 
$\boldsymbol{\theta}=(\sigma_1=0.9,\sigma_2=0.7,\rho=0.65)$. For the shared cause-specific Gamma frailty model, we take $(\alpha^{(1)}_{1},\alpha^{(1)}_{2},\alpha^{(2)}_{1},\alpha^{(2)}_{2}) = (7,6,8.5,10)$ and $\boldsymbol{\theta}=(\sigma_1=0.95,\sigma_2=0.85)$. 
While trying with the correlated cause-specific Gamma frailty model, we encountered some issues with the convergence of likelihood maximization by numerical method. From our experience with real data analysis using this frailty model, we find that the numerical method is more complicated than the previous three models possibly because of larger dimension of the parameter space. Therefore, some manipulation with the initial parameter values is required, which could not be replicated for simulation studies. We intuitively feel that very large and informative datasets are required to produce reasonable estimates for the correlated cause-specific Gamma frailty model. Li et al., (2001) noted that convergence issues of estimators may arise due to irregular behaviour of the log-likelihood function in the neighbourhood of the maxima, which may be the case here as well. \\

For every simulated dataset, we obtain the MLEs of the model parameters and their standard errors using the method described in Section 3. Then, the estimated bias of each parameter estimate is obtained by subtracting the true value of that parameter from the average over the corresponding  estimates from the 10000 simulated samples. For each parameter, the average of the 10000 
standard errors is obtained and denoted by ASE. We also compute the square root of the sample variance  of these $10000$ MLEs to get sample standard error (denoted by SSE). 
Next, for each parameter, we compute the proportion of times the corresponding 10000 approximate $95\%$ confidence intervals based on normal approximation contain the true value. This gives an estimated coverage probability, denoted by CP, of the approximate $95\%$ confidence interval for that paremeter. We carry out the simulation study for sample sizes $n = 50,150,300$. \\

The results for the above three frailty models are summarized in Tables $4.1-4.3$. In each of the three cases, we see decrease in bias, ASE and SSE with increase in sample size, as expected. The values of ASE and SSE tend to be closer with increasing sample size, as expected.  
The CP values also tend to $0.95$ with increasing sample size giving evidence in favour of asymptotic normality. We have carried out similar simulation studies with different values of the model  parameters, but the results are not reported here, as those are qualitatively similar. \\
\newpage
\noindent Table 4.1. Simulation results for shared Gamma frailty model for different joint censoring probability. 
\begin{table}[h!]
    \centering
    \begin{tabular}{|c|c|c|c|c|c|c|c|c|}
\hline
 $p_{cen}$ & $n$ & Properties & $\hat{\alpha}^{(1)}_{1}$ & $\hat{\alpha}^{(1)}_{2}$ & $\hat{\alpha}^{(2)}_{1}$ & $\hat{\alpha}^{(2)}_{2}$ & $\hat{\sigma}$\\
     \hline
  & & Bias & $-0.2310$ & $-0.6091$ & $-0.2798$ & $-0.3960$ & $-0.2281$ \\
  & $50$ & SSE & $1.0839$ & $2.0898$ & $1.5835$ & $1.9169$ & $0.2414$ \\
  & & ASE & $1.3336$ & $2.9576$ & $1.9244$ & $2.4028$ & $0.3601$ \\
  & & CP & $0.8473$ & $0.8545$ & $0.8470$ & $0.8416$ & $0.8977$ \\
  \cline{2-8}
 & & Bias & $-0.1812$ & $-0.4514$ & $-0.2697$ & $-0.3397$ & $-0.1303$\\
$0.1$ & $150$ & SSE & $0.6717$ & $1.4275$ & $0.9280$ & $1.1754$ & $0.1242$ \\
 & & ASE & $0.7963$ & $1.7954$ & $1.1125$ & $1.4081$ & $0.2197$ \\
 & & CP & $0.8943$ & $0.8937$ & $0.8993$ & $0.8995$ & $0.9407$ \\
\cline{2-8}
   & & Bias & $-0.1706$ & $-0.4096$ & $-0.2362$ & $-0.3024$ & $-0.0993$\\
   & $300$ & SSE & $0.4745$ & $1.0294$ & $0.6446$ & $0.8076$ & $0.0842$ \\
   & & ASE & $0.5662$ & $1.2816$ & $0.7948$ & $1.0046$ & $0.1594$ \\
   & & CP & $0.9137$ & $0.9204$ & $0.9216$ & $0.9230$ & $0.9493$\\
\hline
 & & Bias & $-0.1665$ & $-0.4312$ & $-0.1979$ & $-0.2601$ & $-0.2108$ \\
 & $50$ & SSE & $0.9914$ & $1.8407$ & $1.3894$ & $1.6486$ & $0.2257$ \\
 & & ASE & $1.1229$ & $2.3651$ & $1.5632$ & $1.9311$ & $0.3718$ \\
 & & CP & $0.8718$ & $0.8820$ & $0.8780$ & $0.8785$ & $0.9141$ \\
  \cline{2-8}
 & & Bias & $-0.1410$ & $-0.3311$ & $-0.1862$ & $-0.2502$ & $-0.1273$\\
$0.2$ & $150$ & SSE & $0.5840$ & $1.1851$ & $0.8039$ & $0.9666$ & $0.1225$ \\
& & ASE & $0.6612$ & $1.4116$ & $0.9030$ & $1.1206$ & $0.2272$ \\
& & CP & $0.9141$ & $0.9207$ & $0.9130$ & $0.9153$ & $0.9431$ \\
\cline{2-8}
& & Bias & $-0.1210$ & $-0.2986$ & $-0.1790$ & $-0.2322$ & $-0.0893$\\
& $300$ & SSE & $0.4172$ & $0.8445$ & $0.5552$ & $0.6836$ & $0.0848$ \\
& & ASE & $0.4728$ & $1.0082$ & $0.6416$ & $0.7974$ & $0.1645$ \\
& & CP & $0.9284$ & $0.9306$ & $0.9269$ & $0.9271$ & $0.9523$\\
\hline
\end{tabular}
    \label{tab:my_label}
\end{table}

To investigate the effect of joint censoring probability $p_{cen}$ on the different MLEs, we consider two values of $p_{cen}$, 0.1 and 0.2. There is no definite pattern for the bias, ASE, SSE and CP values corresponding to the two values of $p_{cen}$, although the results for $p_{cen}=0.1$ are expected to be `better' than those for $p_{cen}=0.2$. As the joint censoring probability $p_{cen}$ is increased from 0.1 to 0.2, it is found that the probabilities of the other possibilities, including those of one individual being censored, are sometimes  decreased, but still dominating the value of $p_{cen}$. This somehow gives more information on the model parameters, because of some less occurrence of single censoring, giving better properties of the MLEs. \\

\newpage
\noindent Table 4.2. Simulation results for correlated Gamma frailty model for different joint censoring probability. 
\begin{table}[h!]
    \centering
    \begin{tabular}{|c|c|c|c|c|c|c|c|c|c|c|}
\hline
 $p_{cen}$ & $n$ & Properties & $\hat{\alpha}^{(1)}_{1}$ & $\hat{\alpha}^{(1)}_{2}$ & $\hat{\alpha}^{(2)}_{1}$ & $\hat{\alpha}^{(2)}_{2}$ & $\hat{\sigma}_{1}$ & $\hat{\sigma}_{2}$ & $\hat{\rho}$\\
     \hline
 & & Bias & $1.5273$ & $1.0162$ & $1.4087$ & $0.4493$ & $-0.0296$ & $0.0950$ & $-0.0961$  \\
  & $50$ & SSE & $5.1014$ & $3.4022$ & $3.9115$ & $1.3526$ & $0.2782$ & $0.2461$ & $0.3503$\\
 & & ASE & $4.8932$ & $3.2608$ & $3.5827$ & $1.2012$ & $0.3904$ & $0.4091$ & $1.1411$\\
 & & CP & $0.9120$ & $0.9130$ & $0.9801$ & $0.9623$ & $0.9951$ & $0.9926$ & $0.9999$ \\
\cline{2-10}
& & Bias & $0.3964$ & $0.2611$ & $0.3618$ & $0.1146$ & $-0.0246$ & $0.0163$ & $-0.0814$\\
$0.1$ & $150$ & SSE & $2.0332$ & $1.3450$ & $1.2447$ & $0.4102$ & $0.1927$ & $0.1768$ & $0.2864$\\
& & ASE & $1.8929$ & $1.2121$ & $1.1170$ & $0.3839$ & $0.2021$ & $0.2161$ & $0.5177$ \\
& & CP & $0.9176$ & $0.9152$ & $0.9651$ & $0.9595$ & $0.9863$ & $0.9846$ & $0.9994$\\
\cline{2-10}
& & Bias & $0.0234$ & $0.0296$ & $0.1625$ & $0.0507$ & $-0.0145$ & $0.0062$ & $-0.0552$\\
& $300$ & SSE & $1.0641$ & $0.7175$ & $0.6503$ & $0.2284$ & $0.1367$ & $0.1280$ & $0.2244$\\
& & ASE & $1.0192$ & $0.6896$ & $0.6809$ & $0.2365$ & $0.1382$ & $0.1455$ & $0.3051$\\
& & CP & $0.9219$ & $0.9223$ & $0.9661$ & $0.9608$ & $0.9621$ & $0.9833$ & $0.9934$\\
\hline
& & Bias & $1.2595$ & $0.8514$ & $1.1080$ & $0.3538$ & $0.0283$ & $0.1312$ & $-0.0702$\\
 & $50$ & SSE & $4.1206$ & $2.8198$ & $3.0998$ & $1.0841$ & $0.3191$ & $0.2874$ & $0.3315$\\
 & & ASE & $3.7998$ & $2.5914$ & $2.7069$ & $0.9352$ & $0.4383$ & $0.4637$ & $0.9992$\\
 & & CP & $0.9190$ & $0.9126$ & $0.9842$ & $0.9613$ & $0.9944$ & $0.9907$ & $0.9991$\\
\cline{2-10}
& & Bias & $0.2867$ & $0.1916$ & $0.3002$ & $0.0953$ & $-0.0204$ & $0.0336$ & $-0.0589$\\
$0.2$ & $150$ & SSE & $1.4890$ & $1.0073$ & $0.9537$ & $0.3672$ & $0.2196$ & $0.1996$ & $0.2691$\\
& & ASE & $1.3328$ & $0.9109$ & $0.9262$ & $0.3351$ & $0.2347$ & $0.2554$ & $0.4927$\\
& & CP & $0.9271$ & $0.9256$ & $0.9760$ & $0.9606$ & $0.9855$ & $0.9819$ & $0.9954$\\
\cline{2-10}
& & Bias & $0.1141$ & $0.0706$ & $0.1624$ & $0.0498$ & $-0.0165$ & $0.0202$ & $-0.0413$\\
& $300$ & SSE & $0.8716$ & $0.5925$ & $0.5571$ & $0.2065$ & $0.1571$ & $0.1478$ & $0.2096$\\
& & ASE & $0.8411$ & $0.5751$ & $0.5933$ & $0.2157$ & $0.1610$ & $0.1752$ & $0.2914$\\
& & CP & $0.9295$ & $0.9314$ & $0.9717$ & $0.9600$ & $0.9613$ & $0.9776$ & $0.9850$\\
\hline
\end{tabular}
    \label{tab:my_label}
\end{table}
\newpage
\noindent Table 4.3. Simulation results for shared cause-specific Gamma frailty model for different joint censoring probability. 
\begin{table}[h!]
    \centering
\begin{tabular}{|c|c|c|c|c|c|c|c|c|c|}
\hline
 $p_{cen}$ & $n$ & Properties & $\hat{\alpha}^{(1)}_{1}$ & $\hat{\alpha}^{(1)}_{2}$ & $\hat{\alpha}^{(2)}_{1}$ & $\hat{\alpha}^{(2)}_{2}$ & $\hat{\sigma}_1$ & $\hat{\sigma}_2$\\
     \hline
 & & Bias & $0.3132$ & $0.3448$ & $0.4420$ & $0.6502$ & $0.1270$ & $0.1475$\\
 & $50$ & SSE & $3.4688$ & $3.1404$ & $4.0554$ & $4.9451$ & $0.7937$ & $0.9567$\\
 & & ASE & $3.6174$ & $3.3713$ & $4.2811$ & $5.0905$ & $0.8272$ & $0.9276$\\
 & & CP & $0.9170$ & $0.9137$ & $0.9238$ & $0.9205$ & $0.9880$ & $0.9726$\\
  \cline{2-9}
& & Bias & $0.1699$ & $0.1557$ & $0.1678$ & $0.2456$ & $0.0292$ & $0.0364$ \\
$0.1$ & $150$ & SSE & $1.9770$ & $1.8282$ & $2.2922$ & $2.7348$ & $0.4105$ & $0.4756$\\
& & ASE & $1.9402$ & $1.8021$ & $2.2603$ & $2.6988$ & $0.3902$ & $0.4566$\\
& & CP & $0.9388$ & $0.9320$ & $0.9400$ & $0.9402$ & $0.9213$ & $0.9216$\\
\cline{2-9}
& & Bias & $0.0694$ & $0.0762$ & $0.1000$ & $0.1046$ & $0.0120$ & $0.0198$ \\
& $300$ &SSE & $1.3751$ & $1.2829$ & $1.6074$ & $1.9153$ & $0.2807$ & $0.3301$\\
& & ASE & $1.3458$ & $1.2567$ & $1.5767$ & $1.8795$ & $0.2680$ & $0.3166$\\
& & CP & $0.9392$ & $0.9369$ & $0.9385$ & $0.9406$ & $0.9236$ & $0.9286$\\
\hline
 & & Bias & $0.3027$ & $0.3050$ & $0.3074$ & $0.6349$ & $0.1537$ & $0.1482$\\
  & $50$ & SSE & $3.0609$ & $2.7242$ & $3.4885$ & $4.5166$ & $0.8040$ & $0.9393$\\
  & & ASE & $3.1078$ & $2.8411$ & $3.6258$ & $4.4740$ & $0.8195$ & $0.9233$\\
  & & CP & $0.9303$ & $0.9338$ & $0.9381$ & $0.9383$ & $0.9935$ & $0.9790$\\
  \cline{2-9}
& & Bias & $0.1485$ & $0.1033$ & $0.1546$ & $0.2216$ & $0.0271$ & $0.0441$ \\
$0.2$ & $150$ & SSE & $1.6827$ & $1.5469$ & $1.9703$ & $2.4492$ & $0.4022$ & $0.4771$\\
& & ASE & $1.6628$ & $1.5279$ & $1.9442$ & $2.3931$ & $0.3902$ & $0.4680$\\
& & CP & $0.9465$ & $0.9422$ & $0.9454$ & $0.9428$ & $0.9362$ & $0.9306$\\
\cline{2-9}
& & Bias & $0.0659$ & $0.0499$ & $0.0903$ & $0.0952$ & $0.0119$ & $0.0218$ \\
& $300$ & SSE & $1.1745$ & $1.0772$ & $1.3701$ & $1.6915$ & $0.2755$ & $0.3221$\\
& & ASE & $1.1603$ & $1.0715$ & $1.3609$ & $1.6662$ & $0.2701$ & $0.3171$\\
& &CP & $0.9456$ & $0.9451$ & $0.9483$ & $0.9453$ & $0.9358$ & $0.9410$\\
\hline
\end{tabular}
    \label{tab:my_label}
\end{table}

\vskip10pt

\section{Data Analysis}

\vskip10pt

Possibility of hearing loss slowly increases with age, a common phenomenon known as presbycusis. Broadly, there are three causes of hearing loss depending on the part of an ear. \textbf{Sensorineural} hearing loss(SNHL) occurs due to any sensory problem or a neural structure related problem in the inner ear;  \textbf{Conductive} hearing loss is causally related to any hindrance in the outer or middle ear so that sound cannot enter the inner ear properly; and \textbf{Mixed} hearing loss is a combination of both SNHL and Conductive hearing loss. In this section, we analyze the dataset collected from the department of speech and hearing, Ali Yavar Jung National institute of Speech and Hearing Disabilities, Eastern Regional centre, in Kolkata, India. This dataset contains information about hearing loss of $880$ individuals who appeared for diagnosis in the time span of January to February month of the year $2016$. The monitoring time $(X)$ is considered to be the age (in months) of an individual when he(she) appeared for diagnosis. See Koley and Dewanji (2022). 
We intend to use hearing loss information in both the ears so that we have bivariate data. 
We have such information from 796 out of 880 individuals. Since hearing loss can occur due to any of the three above-mentioned causes, this gives a bivariate current status dataset with three competing risks, namely SNHL, Conductive and Mixed with corresponding labels 1, 2 and 3, respectively. A summary of the dataset is presented in Table 5.1 below, with cause of failure 0 meaning censored observation, as mentioned in Section 2. \\

We start by assuming constant (Exponential type) baseline  cause-specific hazards $\alpha^{(k)}_j$'s, denoting the $j$th cause-specific hazard due to cause $j$ for the $k$th individual, for $j = 1,2,3$ and $k = 1,2$. We consider  the four different frailty distribution, as discussed in Section 2. The  MLEs of the model parameters and the  corresponding standard errors (SE) along with the AIC value for each of the four frailty distributions are reported in Tables $5.2$ (for shared and correlated Gamma frailty) and 5.3 (for shared and correlated cause-specific Gamma frailty), respectively. \\

From Tables $5.2$ and 5.3, we observe that  SNHL is the dominating cause of hearing loss in  both ears. It is to be noted that, from Table 5.2, shared and correlated Gamma frailty models give similar results, as evident from the two AIC values. \\

\begin{table}[!h]
\begin{center}
\caption*{Table 5.1. A summary of the hearing loss dataset.}
    \begin{tabular}{|c|c|c|c|c|}
    \hline
    Cause $J_1$ &\multicolumn{4}{|c|}{Cause $J_2$  for $T_2$} \\ \cline{2-5} 
    for $T_1$ & $0$ & $1$ & $2$ & $3$\\
     \hline
    $0$ & $261$ & $0$ & $1$ & $0$\\
     \hline
    $1$ & $0$ & $435$ & $0$ & $7$\\
     \hline
    $2$ & $1$ & $0$ & $28$ & $2$\\
     \hline
    $3$ & $0$ & $7$ & $1$ & $53$\\
     \hline
\end{tabular}
\end{center}
\end{table}

\begin{table}[!h]
\begin{center}
\caption*{Table 5.2. Results for the constant baseline  cause-specific hazard model with shared and correlated Gamma frailty.}
  \begin{tabular}{|c|c|c|c|c|c|c|c|}
\hline
\multicolumn{4}{|c|}{Shared Gamma frailty} & \multicolumn{4}{c|}{Correlated Gamma frailty}\\ 
\hline
Parameter  & MLE & SE & AIC & Parameter  & MLE & SE & AIC\\
  \hline
    $\alpha^{(1)}_{1}$ & $0.8801$ & $0.0840$ & $2274.0813$ & $\alpha^{(1)}_{1}$ & $0.6652$ & $0.0322$ & $2273.3521$\\
     $\alpha^{(1)}_{2}$ & $0.0652$ & $0.0064$ & & $\alpha^{(1)}_{2}$ & $0.0431$ & $0.0022$ &\\
      $\alpha^{(1)}_{3}$ & $0.1281$ & $0.0021$ & & $\alpha^{(1)}_{3}$ & $0.0921$ & $0.0069$ &\\
      $\alpha^{(2)}_{1}$ & $0.8873$ & $0.0847$ & & $\alpha^{(2)}_{1}$ & $0.6392$ & $0.0275$ &\\
      $\alpha^{(2)}_{2}$ & $0.0572$ & $0.0040$ & & $\alpha^{(2)}_{2}$ & $0.0442$ & $0.0212$ &\\
      $\alpha^{(2)}_{3}$ & $0.1314$ & $0.0022$ & & $\alpha^{(2)}_{3}$ & $0.0931$ & $0.0821$ &\\
      $\sigma$ & $6.3803$ & $3.4385$ & & $\sigma_1$ & $2.3903$ & $0.0741$ &\\
      - & - & - & & $\sigma_2$ & $2.3993$ & $0.0782$ &\\
      - & - & - & & $\rho$ & $0.9940$ & $0.0170$ &\\
     \hline
\end{tabular}  
\end{center}
\end{table}

From Table $5.2$, we note that the magnitudes of the m.l.e's of $\sigma_1$ and $\sigma_2$ for constant cause-specific baseline hazard with correlated Gamma frailty are very close. Therefore, it is of interest to test the hypothesis, $H_{0}:\sigma_{1} = \sigma_{2}$ vs $H_{1}:\sigma_{1} \neq \sigma_{2}$. The likelihood ratio test statistic comes out to be $2.26$, less than $3.84$, the upper $0.05$ point of $\chi^{2}_{1}$ distribution. Hence, we fail to reject the null hypothesis of equality of frailty variances at $0.05$ level of significance.  

We next fit the two cause-specific frailty models with constant cause-specific hazards. For the shared cause-specific Gamma frailty model, we have three frailty parameters $\sigma_1,\sigma_2$ and $\sigma_3$. For the correlated cause-specific Gamma frailty model, we have nine frailty parameters $\sigma_{kj},\ \rho_j$, for $j=1,2,3$ and $k=1,2$. In view of the results of Table 5.2, it is of interest to test for the equality of $\sigma_{1j}$ and $\sigma_{2j}$, for $j=1,2,3$, so that there may be some reduction in number of parameters. The likelihood ratio test for this hypothesis, for the sample at hand, leads to the observed value of the test statistic as $0.9996$ to be compared with $\chi^{2}_3$ distribution. This is less than the corresponding 95th percentile.  Therefore, we fail to reject the equality at $5\%$ significance level. So, we assume this equality $\sigma_{1j}=\sigma_{2j}=\sigma_j$ (say), for $j=1,2,3$, for further analysis and the results are reported in Table 5.3. \\
 
\begin{table}[!h]
\begin{center}
\caption*{Table 5.3. Results for the constant  baseline  cause-specific hazard model with shared and correlated cause-specific Gamma frailty.}
  \begin{tabular}{|c|c|c|c|c|c|c|c|}
\hline
\multicolumn{4}{c}{Shared cause-specific Gamma frailty} & \multicolumn{4}{c}{Correlated 
cause-specific Gamma frailty}\\  
\hline
Parameter  &  MLE & SE & AIC & Parameter  &  MLE & SE & AIC \\ \hline
    $\alpha^{(1)}_{1}$ & $35.95820$ & $31.16312$ & $1838.96400$ & $\alpha^{(1)}_{1}$ & $2485.23120$ & $75.07471$ & $1773.84080$\\
     $\alpha^{(1)}_{2}$ & $0.00605$ & $0.00491$ & & $\alpha^{(1)}_{2}$ & $0.02021$ & $0.00767$ &\\
      $\alpha^{(1)}_{3}$ & $0.00338$ & $0.00102$ & & $\alpha^{(1)}_{3}$ & $0.01821$ & $0.01634$ &\\
      $\alpha^{(2)}_{1}$ & $39.21070$ & $37.42933$ & & $\alpha^{(2)}_{1}$ & $2645.33701$ & $10.79667$ &\\
      $\alpha^{(2)}_{2}$ & $0.00550$ & $0.00431$ & & $\alpha^{(2)}_{2}$ & $0.01867$ & $0.01842$ &\\
      $\alpha^{(2)}_{3}$ & $0.00345$ & $0.00101$ & & $\alpha^{(2)}_{3}$ & $0.01881$ & $0.01402$ &\\
      $\sigma_1$ & $12.92450$ & $1.36541$ & & $\sigma_1$ & $4.65674$ & $0.93314$ &\\
      $\sigma_2$ & $45.58919$ & $18.59252$ & & $\sigma_2$ & $5.07337$ & $4.21343$ &\\
      $\sigma_3$ & $7.65704$ & $2.66357$ & & $\sigma_3$ & $1.99483$ & $2.47974$ &\\
      - & - & - & & $\rho_1$ & $0.99900$ & $1.06120$ &\\
      - & - & - & & $\rho_2$ & $0.99950$ & $1.03113$ &\\
      - & - & - & & $\rho_3$ & $0.99930$ & $1.01319$ &\\
     \hline
\end{tabular}  
\end{center}
\end{table}

Note that the AIC value for the correlated cause-specific Gamma frailty is the lowest indicating the best fit with constant baseline cause-specific hazards. For this model, the parameter $\sigma_1$ appear to be significant. So, there is evidence of dependence between occurrence of hearing loss in left and right ears due to the cause SNHL, which seems to be the most dominating cause. \\

Next, we consider a Weibull type baseline cause-specific hazards (that is, power functions of time) with the scale parameters $\alpha^{(k)}_j$ and a common shape parameter $\gamma^{(k)}$, for $j = 1,2,3$ and $k = 1,2$. We use the four Gamma frailty models as before. The results for shared Gamma frailty and correlated Gamma frailty models are presented in Table 5.4. Note that the AIC values are decreased compared to those for the constant cause-specific hazards in Table 5.2. There is evidence of time dependence on the  baseline cause-specific hazards from the estimates of $\gamma^{(1)}$ and $\gamma^{(2)}$, indicating increasing cause-specific hazards with age. Both the shared and correlated Gamma frailty models give similar results, as for the constant baseline cause-specific hazards. \\

\begin{table}
\begin{center}
\caption*{Table 5.4. Results for the power function  baseline cause-specific hazard model with shared and correlated Gamma frailty.}
  \begin{tabular}{|c|c|c|c|c|c|c|c|} \hline
\multicolumn{4}{c}{Shared Gamma frailty} & \multicolumn{4}{c}{Correlated Gamma frailty}\\  
\hline
Parameter  &  MLE & SE & AIC & Parameter  &  MLE & SE($\times 100$) & AIC\\
  \hline
    $\alpha^{(1)}_{1}$ & $0.0387$ & $0.0031$ & $2048.8942$ & $\alpha^{(1)}_{1}$ & $0.0380$ & $0.7342$ & $2032.12$\\
     $\alpha^{(1)}_{2}$ & $0.0317$ & $0.0035$ & & $\alpha^{(1)}_{2}$ & $0.0316$ & $0.0673$ &\\
      $\alpha^{(1)}_{3}$ & $0.0334$ & $0.0034$ & & $\alpha^{(1)}_{3}$ & $0.0331$ & $0.5314$ &\\
      $\alpha^{(2)}_{1}$ & $0.0421$ & $0.0042$ & & $\alpha^{(2)}_{1}$ & $0.0378$ & $0.7243$ &\\
      $\alpha^{(2)}_{2}$ & $0.0344$ & $0.0042$ & & $\alpha^{(2)}_{2}$ & $0.0314$ & $0.7482$ &\\
      $\alpha^{(2)}_{3}$ & $0.0361$ & $0.0041$ & & $\alpha^{(2)}_{3}$ & $0.0329$ & $0.8034$ &\\
      $\gamma^{(1)}$ & $13.5943$ & $2.7454$ & & $\gamma^{(1)}$ & $14.2930$ & $11.4774$ &\\
      $\gamma^{(2)}$ & $12.3653$ & $2.5281$ & & $\gamma^{(2)}$ & $14.3522$ & $3.1196$ &\\
      $\sigma$ & $18.3432$ & $3.9534$ & & $\sigma_{1}$ & $4.4499$ & $1.1413$ &\\
      - & - & - & & $\sigma_{2}$ & $4.4233$ & $1.1714$ &\\
      - & - & - & & $\rho$ & $0.9900$ & $0.7224$ &\\
     \hline
\end{tabular}  
\end{center}
\end{table}
\begin{table}[!h]
\begin{center}
\caption*{Table 5.5. Results for the power function  baseline cause-specific hazard model with shared and correlated cause-specific Gamma frailty.}
  \begin{tabular}{|c|c|c|c|c|c|c|c|} \hline
\multicolumn{4}{c}{Shared cause-specific Gamma frailty} & \multicolumn{4}{c}{Correlated cause-specific Gamma frailty}\\  
\hline
Parameter  &  MLE & SE & AIC & Parameter  &  MLE & SE & AIC\\
  \hline
    $\alpha^{(1)}_{1}$ & $35.4486$ & $5.0221$ & $1671.86000$ & $\alpha^{(1)}_{1}$ & $35.88337$ & $4.79201$ & $137.00446$\\
     $\alpha^{(1)}_{2}$ & $0.00813$ & $0.00186$ & & $\alpha^{(1)}_{2}$ & $30.88353$ & $3.20406$ &\\
      $\alpha^{(1)}_{3}$ & $0.00493$ & $0.00152$ & & $\alpha^{(1)}_{3}$ & $0.25663$ & $0.95197$ &\\
      $\alpha^{(2)}_{1}$ & $38.77887$ & $4.84707$ & & $\alpha^{(2)}_{1}$ & $41.44030$ & $0.87682$ &\\
      $\alpha^{(2)}_{2}$ & $0.00782$ & $0.00180$ & & $\alpha^{(2)}_{2}$ & $19.31556$ & $12.28719$ &\\
      $\alpha^{(2)}_{3}$ & $0.00500$ & $0.00158$ & & $\alpha^{(2)}_{3}$ & $2.75814$ & $0.20759$ &\\
      $\gamma^{(1)}$ & $2.92518$ & $0.38221$ & & $\gamma^{(1)}$ & $69.67167$ & $0.66677$ &\\
      $\gamma^{(2)}$ & $2.89409$ & $0.37892$ & & $\gamma^{(2)}$ & $63.78004$ & $0.18248$ &\\
      $\sigma_1$ & $31.09130$ & $4.26321$ & & $\sigma_1$ & $21.26987$ & $1.88782$ &\\
      $\sigma_2$ & $37.83996$ & $4.11579$ & & $\sigma_2$ & $88.66456$ & $10.32503$ &\\
      $\sigma_3$ & $5.08912$ & $4.15168$ & & $\sigma_3$ & $39.65239$ & $3.18067$ &\\
      - & - & - & & $\rho_1$ & $0.99010$ & $1.06647$ &\\
      - & - & - & & $\rho_2$ & $0.99020$ & $0.97282$ &\\
      - & - & - & & $\rho_3$ & $0.93218$ & $1.17004$ &\\
     \hline
\end{tabular}  
\end{center}
\end{table}
\newpage
Similar to the case of constant cause-specific hazard, we test the equality of frailty variances (that is, $\sigma^2_{1} = \sigma^2_{2}$). In this case, the likelihood ratio test statistic comes out to be $2.469$. So, we cannot reject the null hypothesis since this value is less than $3.84$. \\
Finally, we consider the two cause-specific frailty models with power function (that is, Weibull type)  baseline cause-specific hazards having common shape parameter for the three causes. As before, we assume  $\sigma_{1j}=\sigma_{2j}=\sigma_j$, for $j=1,2,3$. The results are presented in Table 5.5. We observe better fits compared to all other models in terms of AIC value (See Tables 5.2-5.4).  
However, the large values of the estimates of $\gamma^{(1)}$ and $\gamma^{(2)}$, indicating strong time dependence, are somewhat unusual. One may wonder if this is a case of some over-fitting. Therefore, although the model with Weibull type baseline cause-specific hazards and correlated cause-specific frailty gives the least value of AIC, one should use caution before adopting this model. In summary, there is evidence of strong time dependence on the different cause-specific hazards with the cause SNHL dominating other causes. Also, there is evidence of some association between occurrence of hearing loss in left and right ears, the nature of which seems to depend on the underlying cause. \\

\vskip10pt

\section{Concluding Remarks}

\vskip10pt

In this work, we consider parametric modeling and analysis of bivariate current status data with competing risks, while association  between the two individuals in a pair is modeled through four different Gamma frailty distributions possibly depending on the cause of failure. These four different frailty distributions cover a whole range of dependence as discussed in Section 2. The baseline cause-specific hazards are assumed to belong to a parametric family that includes many commonly used parametric hazard functions. However, there are some other parametric hazard functions, for example, Log-normal, Gompertz, Pareto, generalized Gamma, etc., which do not belong to this family. A remaining task is to construct and analyze models consisting of the above-mentioned cause-specific hazard functions with the four different Gamma frailty distributions. \\

Note that we have used different Gamma frailty in this work because of its popularity and mathematical tractability. It is also an important task to consider other frailty distributions, for example, inverse Gaussian, Log-normal (See Gorfine and Hsu, $2011$), etc., which also produce tractable expressions in many cases, and consequently analyze the resulting models for bivariate current status data with competing risks. From our data analysis in Section 5, it is seen that the fit of the model substantially depends on the kind of frailty distribution. In view of that, it is of interest to investigate different types of frailty distributions subject to identifiability of the resulting bivariate model. This applies to routine analysis of bivariate failure time data as well using some specific frailty distribution. \\

As evident from Sections 4-5, our analysis involves heavy computation and some manipulation with initial values because of some irregular behaviour of the likelihood in the neighbourhood of the true maxima. This numerical problem gets further compounded with larger number of competing risks, as is evident from our experience with two causes in the simulation study and three causes in the data analysis. Even if the model is identifiable, one needs large number of observations to tackle this problem of numerical instability. With a smaller dataset, one may be advised to consider a simpler frailty model and possibly combine several causes to have fewer number of competing risks. \\

Extending this work to multivariate current status data with competing risks seems, in principle, straightforward. But it requires quite an amount of complications in terms of notation, mathematical tractibility to derive identifiability results, etc., and also the computational issues for analysis.
\section{References}
Ansa, A. A., and Sankaran, P. G. (2005). Estimation of bivariate survivor function of competing risk models under censoring. J Stat Theory Appl, 4, 404-423.\\
Bandeen‐Roche, K., and Liang, K. Y. (2002). Modelling multivariate failure time associations in the presence of a competing risk. Biometrika, 89(2), 299-314.\\
Cui, Q., Zhao, H., and Sun, J. (2018). A new copula model-based method for regression analysis of dependent current status data. Statistics and Its Interface, 11(3), 463-471.\\
Dabrowska, D. M. (1988). Kaplan-Meier estimate on the plane. The Annals of Statistics, 1475-1489.\\
Duchateau, L., and Janssen, P. (2008). The frailty model. New York: Springer Verlag.\\
Gorfine, M., and Hsu, L. (2011). Frailty-based competing risks model for multivariate survival data. Biometrics, 67(2), 415-426.\\
Heckman, J., and Singer, B. (1984). The identifiability of the proportional hazard model. The Review of Economic Studies, 51(2), 231-241.\\
Hu, T., Zhou, Q., and Sun, J. (2017). Regression analysis of bivariate current status data under the proportional hazards model. Canadian Journal of Statistics, 45(4), 410-424.\\
Jewell, N. P., Van Der Laan, M., and Lei, X. (2005). Bivariate current status data with univariate monitoring times. Biometrika, 92(4), 847-862.\\
Johnson, N. L., and Kotz, S. (1975). A vector multivariate hazard rate. Journal of Multivariate Analysis, 5(1), 53-66.\\
Koley, T., and Dewanji, A. (2022). Current status data with two competing risks and missing failure types: a parametric approach. Journal of Applied Statistics, 49(7), 1769-1783.\\
Ma, L., Hu, T., and Sun, J. (2015). Sieve maximum likelihood regression analysis of dependent current status data. Biometrika, 102(3), 731-738.\\
Oakes, D. (1989). Bivariate survival models induced by frailties. Journal of the American Statistical Association, 84(406), 487-493.\\
Prenen, L., Braekers, R., and Duchateau, L. (2017). Extending the Archimedean copula methodology to model multivariate survival data grouped in clusters of variable size. Journal of the Royal Statistical Society Series B: Statistical Methodology, 79(2), 483-505.\\
Sun, T., and Ding, Y. (2021). Copula-based semiparametric regression method for bivariate data under general interval censoring. Biostatistics, 22(2), 315-330.\\
Wang, L., Sun, J., and Tong, X. (2008). Efficient estimation for the proportional hazards model with bivariate current status data. Lifetime Data Analysis, 14, 134-153.\\
Wienke, A. (2010). Frailty models in survival analysis. Chapman and Hall/CRC.\\
Yashin, A. I., Vaupel, J. W., and Iachine, I. A. (1995). Correlated individual frailty: an advantageous approach to survival analysis of bivariate data. Mathematical population studies, 5(2), 145-159.
\end{document}